\documentclass[a4paper,12pt]{article}

\usepackage{graphicx}
\usepackage{amstext}
\usepackage[]{ntheorem}
\usepackage{fancyhdr}
\usepackage{amsfonts}
\usepackage{amssymb}
\usepackage{amsmath,chemarrow}
\usepackage{array}
\usepackage{multirow}

\usepackage{t1enc}
\usepackage{tikz}
\usepackage[utf8]{inputenc}
\usepackage{authblk}
\usepackage{hyperref}
\title{Convex Combinatorial Auction of Pipeline Network Capacities}
\author[1]{Dávid Csercsik}
\affil[1]{\small{Faculty of Information Technology and Bionincs, P\'{a}zm\'{a}ny P\'{e}ter Catholic University, 1083 Práter u. 50/A, Budapest, Hungary}}

\begin{document}

\maketitle

\abstract{In this paper we propose a mechanism for the allocation of pipeline capacities, assuming that the participants bidding for capacities do have subjective evaluation of various network routes. The proposed mechanism is based on the concept of bidding for route-quantity pairs. Each participant defines a limited number of routes and places multiple bids, corresponding to various quantities, on each of these routes. The proposed mechanism assigns a convex combination of the submitted bids to each participant, thus its called convex combinatorial auction. The capacity payments in the proposed model are determined according to the Vickrey-Clarke-Groves principle. We compare the efficiency of the proposed algorithm with a simplified model of the method currently used for pipeline capacity allocation in the EU (simultaneous ascending clock auction of pipeline capacities) via simulation, according to various measures, such as resulting utility of players, utilization of network capacities, total income of the auctioneer and fairness.}

\section{Introduction}
%
%
\label{Motivation}

The European natural gas network represents an enormous infrastructure system, which is also constantly in the focus of geopolitics \cite{bilgin2009geopolitics,ericson2009eurasian}. In the traditional model, national or multinational energy companies built their own pipelines requiring huge investments, and expected that their latter trade transactions using the pipeline will provide them with sufficient returns.
Nowadays, exclusive ownership is not the general institutional setting. Many pipelines within the European Union are subject to regulated third party access (TPA). Since the early 1990s, the EU have adopted a number of increasingly assertive directives and regulations to develop the common market for gas by ensuring fair TPA access to the transportation system within the Union --- see \cite{eu1991directive,eu1998directive,eu2003directive,eu-1775-2005,EU2009reg}. According to this scheme, the member countries have established a system of transport fees overseen by a regulatory authority. Under such a regime, the owner of a pipeline no longer enjoys exclusive right over the transport capacities. Instead, he has to grant access, provided he is compensated according to the regulated tariff.
Cooperative game theoretic analysis of TPA and the implied transfer profits in natural gas networks has been proposed in \cite{CSERCSIK_2019EE}.

\subsection{Motivation: Current practice of pipeline network capacity allocation in the EU}

A decade ago, regulations of the third energy package \cite{EU2009directive_gas} basically separated the network operation from the trading and supply, expanded the rights of regulation authorities, and created the Agency for the Cooperation of Energy Regulators (ACER), and the European Network of Transmission System Operators for Gas (ENTSOG). As a result, in the last 10 years, the bias of trading already significantly shifted from long term (usually fixed-price) contracts to more liquid trading platforms (markets corresponding to so called \emph{gas hubs}). In this framework, to provide infrastructure for such increasingly interactive trading, the transmission system operators (TSOs) market the transfer capacities of pipelines as standardized products of variable time-frames (from yearly to intra-day intervals). According to the reports of \cite{ACER2016}, the volume of engagements corresponding to short-time products constantly increases.
The efficiency of capacity bookings in Germany has been discussed by \cite{keller2019mergers}.

As long as the capacities required for the planned trade transactions do not exceed the pipeline capacities, allocation is simple, and it practically means only administration. However if the available capacities are not enough to satisfy all participants aiming to allocate capacities in the network,
some kind of capacity-allocation method must be used to distribute the available pipeline capacities among
participants (players) who apply for them. The first auction, which coordinated the long-term bookings of available existing and future pipeline capacities on the EU-level has been held in 2017 March on the PRISMA auction platform.
During this auction, yearly, quarterly and monthly pipeline capacity products have been
auctioned simultaneously using an ascending clock auction (ACA).
Altogether 2165 unique auctions took place on 6 March for each point and each year. As pointed out by \cite{REKK2017}, in most of the cases no real competition emerged, and as the result of this auction, the dominant market player (GAZPROM) was able to acquire the great majority of high-importance capacity licenses for in some cases as long as 20 years (for example, \emph{all} interconnection capacities on the border of Slovakia have been booked for 20-25 years by GAZPROM).

Several factors may be identified as underlying causes for this result. First, Russia, unlike other suppliers of Europe, like Algeria or Norway, typically delivers gas to the border of the importer country, thus countries which import gas from Russia do have modest interest in acquiring transport routes. The reasons for this are partially historical -- in deals of the former decades, the market power of GAZPROM was very high, so importers were compelled to agree with such details of bargains.

Second, if a large producer supplying a significant number of clients aims to buy capacities for his deliveries, all the delivery paths in question originate from the production site, and they have potentially large overlaps (see e.g. the Nord Stream I and II and their connected pipelines, which will practically supply the majority of Europe). In this case, it is easy to identify pipelines and interconnection points which are critical for these delivery projects, and thus represent high value for the player. In other words, the optimal bidding strategy of such large producers is quite straightforward in the current framework, while they typically also have the resources to obtain capacity licenses for long periods.

In contrast, the optimal bidding strategy in the current framework is not trivial for smaller consumers.
Consumers typically buy gas the on established hubs, the prices of which may be different and also uncertain regarding longer periods (e.g. years) for which capacity may be booked in the present system.
Such hubs correspond to \emph{market areas}, inside which the physical transportation of gas is the responsibility of the TSO. The article of \cite{keller2019mergers} discusses the implications of these market areas in Germany, and analyzes the efficiency of inter-area capacity bookings.

Overall, it can be said that the current allocation practice and the respective algorithms do have their pros and cons, but in general it is reasonable to ask if there's any alternative to the current method of capacity allocation. In the current paper we propose exactly such an alternative approach, called convex combinatorial auction (CCA) of pipeline capacities. As a first step, we define and test this method on an abstract model under several simplifying assumptions (see subsection \ref{subsec_assumptions}).

We consider a scenario where, under the principle of regulated third party access, local (national) TSOs have the right to determine transfer fees for their pipelines, but the pipeline capacity licenses are allocated by a central authority via auction. We compare the newly proposed CCA-based allocation to the allocation based on the simultaneous ACA, assuming a simple but reasonable optimal bidding strategy of the participants of the ACA (see subsection \ref{optimal_bidding_ACA}). We use various measures for the comparison, such as resulting utility of players, utilization efficiency of network infrastructure, total amount of payment for the capacity rights (income of the auctioneer) and fairness. The structure of the paper is as follows. In section \ref{section_MM} we define the principles of the used model of the network, and the algorithms modelling the ACA and CCA based allocations, and demonstrate the concepts on a simple, small example. In section \ref{sec_results}, we perform simulations on high numbers of randomized scenarios to get statistical data about the performance of the two methods and summarize the respective results.
Section \ref{sec_discussion} discusses the results of the simulations and the properties of the two different allocation mechanisms, while section \ref{sec_conclusions} concludes.

\section{Materials and methods}
\label{section_MM}
In the following, an example with a simple network will be introduced to
demonstrate the modelling and simulation assumptions corresponding to the two allocation methods considered.

After defining the network and related parameters, we will first discuss the simulation of the ACA for the allocation of pipeline capacities, and determine the resulting capacity rights, payments and utilities in this framework, which is motivated by the current practice of the EU for capacity allocation \cite{EU2017-CommRegulation}.
Following this, we introduce the proposed alternative method, the CCA, demonstrate how it can be applied to the proposed example, and compare the results with the outcome of the ACA.

In this proposed simple example, we will focus on consumers, and we will assume that they are the only participants of the capacity auction. Regarding realistic scenarios, at least in addition to local gas distribution companies, who may be considered as consumers, multinational energy companies and traders are also present as typical bidders of such auctions.
The benefit of considering only consumers as bidders is however that, as we will see, according to their explicit demand elasticity characterization, their rational bidding strategy (under a few additional assumptions in the case of ACA) may be easily derived -- this task would be much more harder in the case of agents representing multinational companies with more complex incentives.

In the current work we focus on long-term capacity rights.
The regulation \cite{EU2017-CommRegulation} defines yearly, quarterly, monthly, daily and within-day capacity products, from which the first three are sold via the ACA algorithm. In other words this
means that if one is willing to allocate capacities for the first month of the year, he/she has 3 opportunities to do it.

Motivated by this, our model of the ACA process will have three steps. In the first step we assume that
all capacities of the network in question are subject to auction. After the first step has finished, remaining (not-allocated capacities) are subject to the second round of ACA auctions, and so similarly, the remaining not-allocated capacities are subject to the last round of ACA auctions\footnote{Let us note that according to the current practice, 10 \% of the available transfer capacity is reserved for short term trading.}.


%

\subsection{Modelling assumptions and the example network}
\label{subsec_assumptions}

\subsubsection{Market areas and their representation in the model}

As discussed by \cite{keller2019mergers}, market areas (MAs) are sets of physical network nodes, between which the transportation of gas is the responsibility of the TSOs. Network users are able to inject gas at any entry point of the MA and withdraw gas at any exit point that belongs to the same MA, if they have the capacity rights for the respective entry and exit points.
Some entry points of a MA may correspond to production sites, while others may represent incoming pipelines. Let us note, that multiple such pipelines exists (see Fig. 3 in \cite{keller2019mergers}).

In the terms of our model, the nodes represent MAs, and the edges represent the capacities connecting them. According to the above considerations, it is possible that more than one edge is present between two nodes. Although we do not consider such cases in the paper, the model is capable of handling these scenarios.

\subsubsection{Bundled products and their representation in the model}

Similar to transfer fees, capacity products in the practical European system (PRISMA - see \url{https://www.prisma-capacity.eu/}) are also considered corresponding to the entry and exit points of MAs. The basic reason for this is that the transfer capacities are managed locally by the TSO's of the respective price zone. To make the life of bidders easier, the capacity allocation platform defines so called bundled products, composed of an exit and an entry capacity. This means that if I'd like to transfer from node A to node B, I have the possibility to bid for a bundled AB product, in which the exit capacity of A and the entry capacity of B are included. These bundled products are handled in the PRISMA system in a way, which ensures that the total entry and exit capacities of price zones are respected.

For the aim of simplicity, in the used modelling framework we consider only such bundled capacity products. Let us note however, that the used methodology may be easily generalized to a more detailed scenario.
The current model takes capacity from A to B into account as a product, if node A is connected to node B.
If one would like to consider entry and exit capacities distinctly, an intermedier node X may be introduced on the edge A-B. In this case the edge A-X represents the entry and exit capacities of A, while the edge X-B represents the entry and exit capacities of zone B.

\subsubsection{The example network}

\paragraph*{Modelling the sources and source costs}

In the proposed model framework it is assumed that natural gas is available at distinguished nodes (representing market areas), from where consumers must ensure themselves routes to transport it to consumption sites. In the current model we assume that these sources are able to provide arbitrary quantities on prices, which are fixed for the period in question (for which we consider the allocation of transfer capacities). Let us note that in the case of realistic scenarios, the price of natural gas at the trading hubs may be significantly volatile, and depends on the nature of the source as well (obtained e.g. from actual transports, gas reservoirs or from LNG terminals).
According with the recent line of EU regulations, we assume that no price differentiation is allowed at the market.

\begin{figure}[h!]
  \centering
  \includegraphics[width=8.5cm]{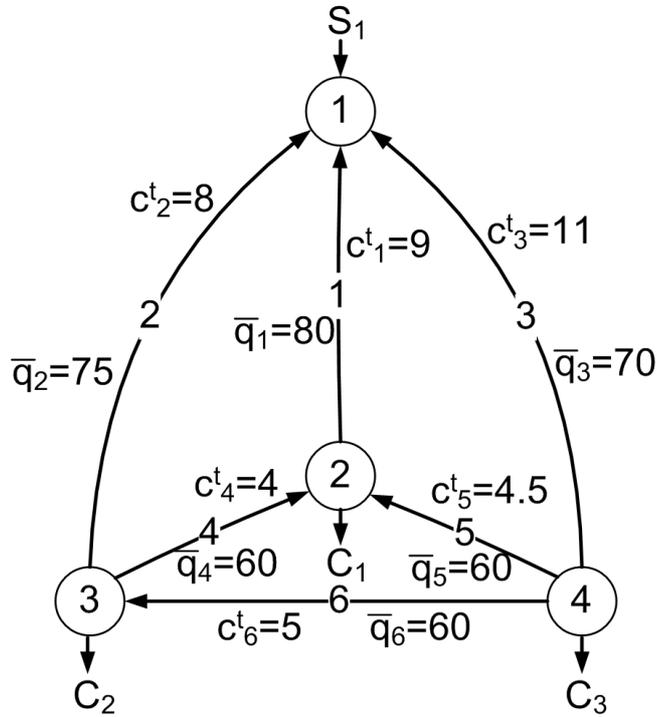}\\
  \caption{Example network. Each node and edge is labeled with its ID. $S_1$ denotes the only source present in this simple example network, while $C_1$, $C_2$, and $C_3$ denote the consumers. $c^t_i$ denotes the transfer cost on line $i$, while $\overline{q}_i$ denotes the maximal capacity of the line.}\label{NCA_example_1}
\end{figure}

Let us consider the network depicted in Fig. \ref{NCA_example_1}.
Based on the above considerations, the source $S_1$ in the model (at node 1) provides gas for every consumer at the same price (23 EUR/MMBtu -- EUR per Million Metric British unit)

We assume one source ($S_1$) located in node 1, and three consumers ($C_1-C_3$) located in nodes 2, 3 and 4 respectively. As in the current study we focus on network capacity allocation, we assume that the sources can provide arbitrary quantities (we assume no inlet limits).

We assign a direction to each edge, as denoted in the Figure, to account for positive and negative flows, but we assume that the pipelines corresponding to the edges are bi-directional, and their maximal transfer capacity ($\overline{q}_j$ for pipeline $j$) is the same in both directions.
We also assume that the transfer cost for pipeline ($c^t_j$ for pipeline $j$) is also the same in the positive and in the negative direction. In practice, local TSOs set entry and exit fees at interconnection points, but from these values transfer fees of a certain line $i$ in the context of the model may be easily derived as the exit fee of the source point and entry fee of the destination point.

\begin{table}[h!]
\begin{center}
\begin{tabular}{|c|c|c|}
  \hline
    & $\bar{q}$ & $c^t$  \\ \hline
  1 & 80        & 9     \\
  2 & 75        & 8     \\
  3 & 70        & 11    \\
  4 & 60        & 4     \\
  5 & 60        & 4.5   \\
  6 & 60        & 5     \\
  \hline
\end{tabular}
\caption{Edge parameters in the network of example 1: Maximal capacity ($\bar{q}$) and transfer cost ($c^t$). \label{Table_edge_par_EX1}}
\end{center}
\end{table}

\begin{figure}[h!]
  \centering
  \includegraphics[width=5cm]{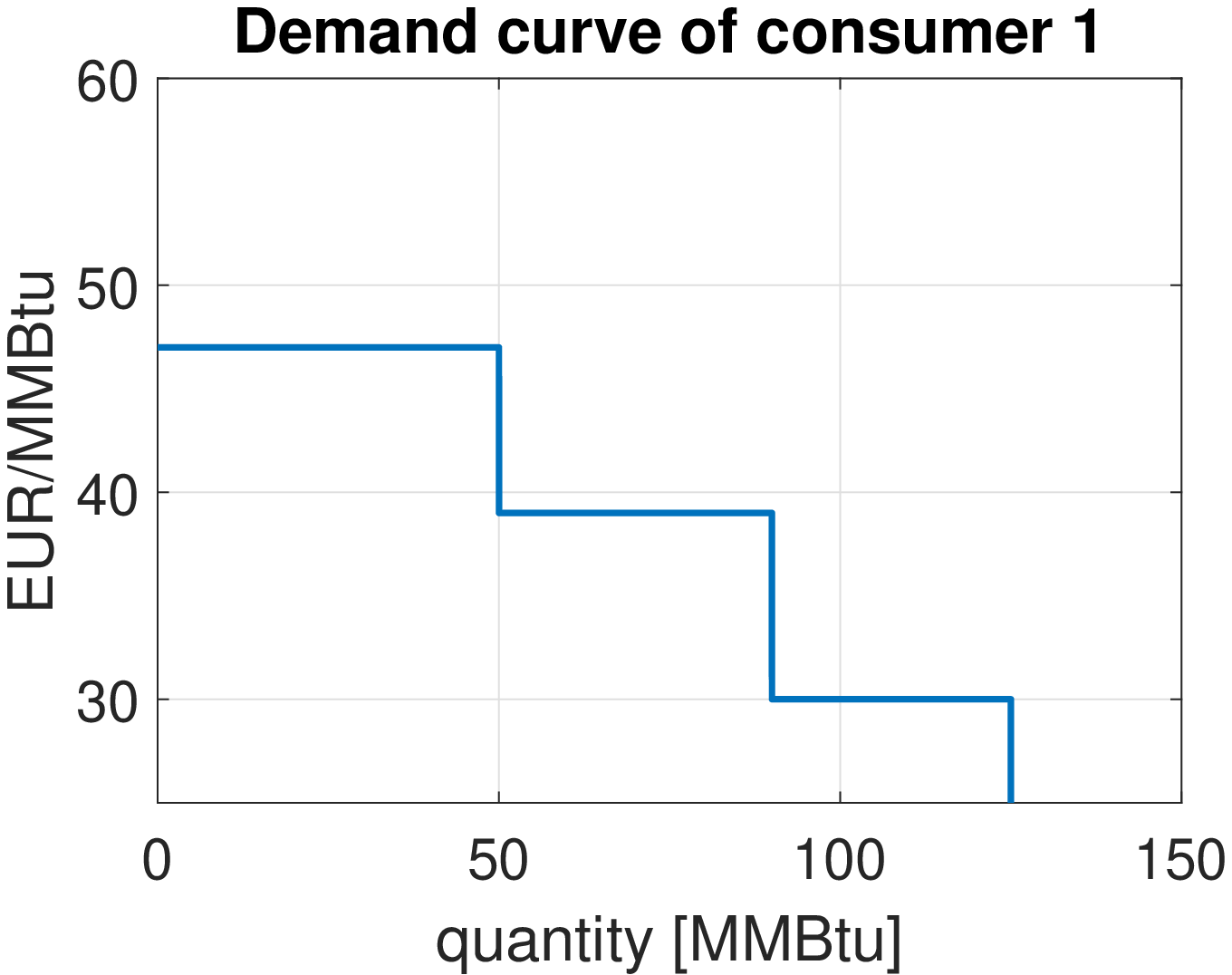}~
  \includegraphics[width=5cm]{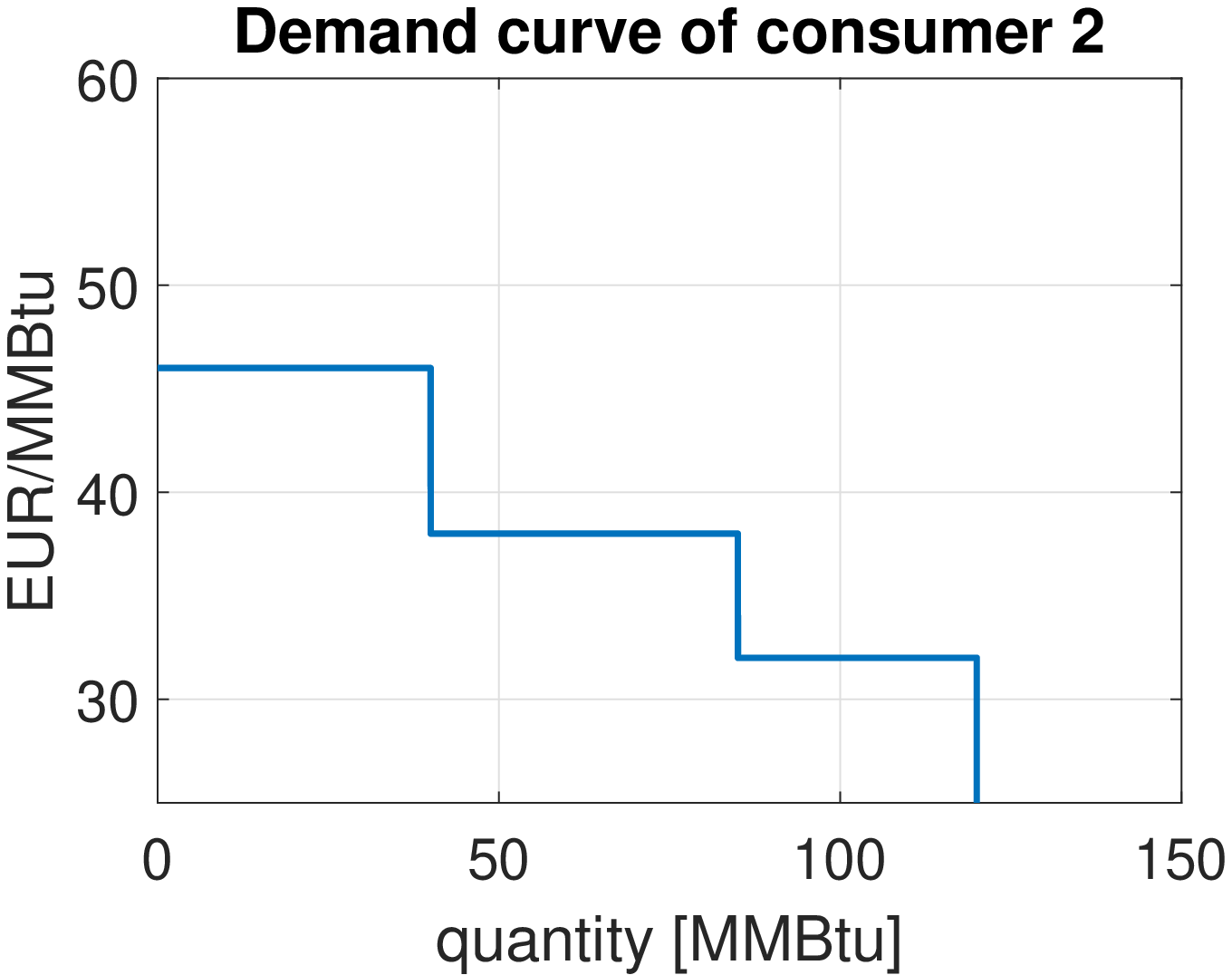}
  \includegraphics[width=5cm]{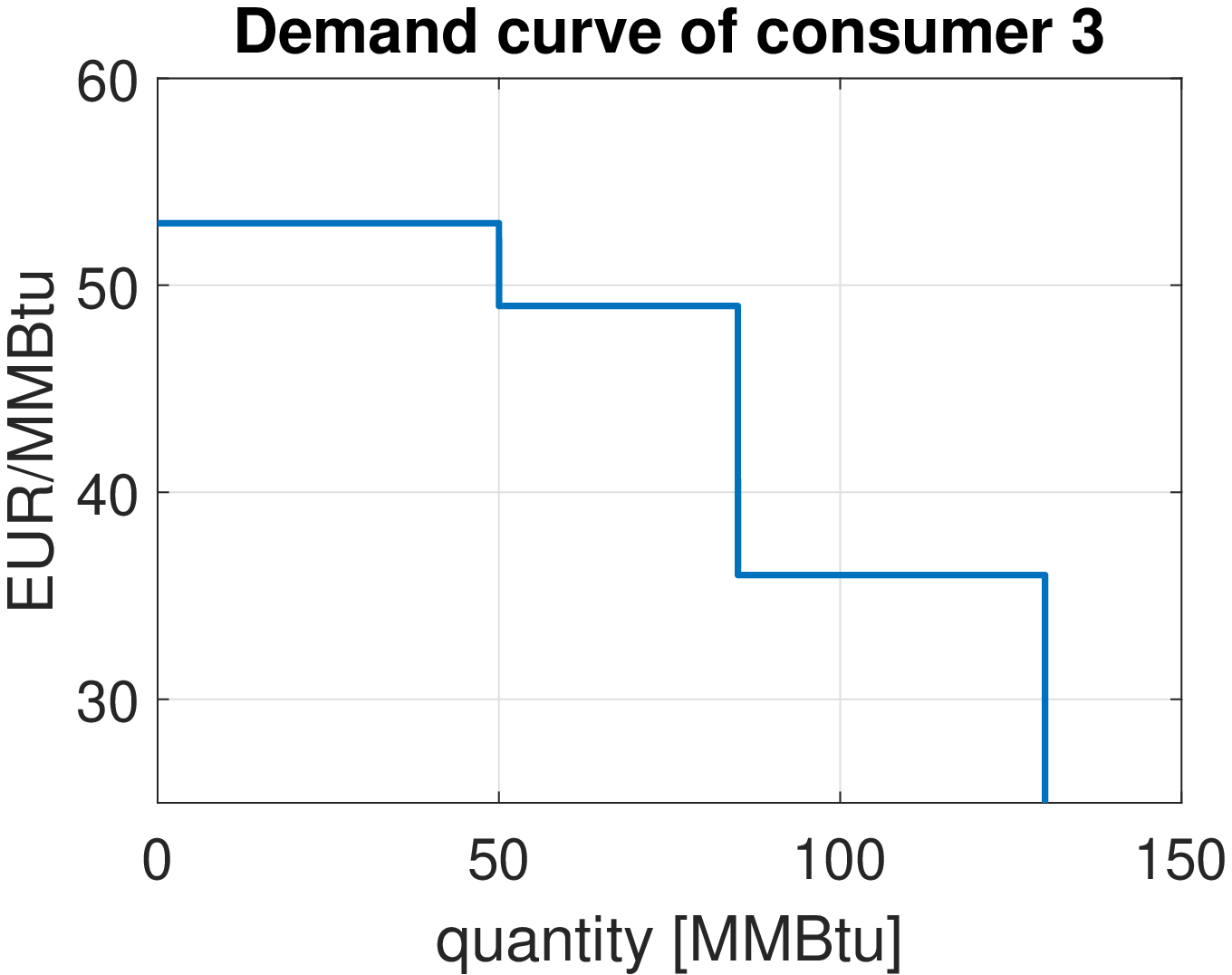}
  \caption{Inverse demand functions of consumers assumed in Example 1. MMBtu stands for million metric British unit.}\label{NCA_example_DC_1}
\end{figure}

\paragraph*{Consumer demand}
\label{consumer_demand}
We use piecewise constant inverse demand curves for the description of demand elasticity as depicted in Fig. \ref{NCA_example_DC_1}. Each piecewise constant part has two parameters: A price ($P$) and a consumption quantity ($Q$). In this formalism $P^i_j$ denotes the price level of the $j$-th step of the inverse demand function of player $i$. The parameters of the demand functions used in the example and depicted ion Fig. \ref{NCA_example_DC_1} are summarized in Table \ref{Table_IDF}.

\begin{table}[h!]
\begin{small}
\begin{center}
\begin{tabular}{|c|c|c|c|c|c||c|c|c|c|c|c|}
  \hline
  $P^1_1$ & 47 & $P^2_1$ & 46 & $P^3_1$ & 53 & $Q^1_1$ & 50 & $Q^2_1$ & 40 & $Q^3_1$ & 50 \\ \hline
  $P^1_2$ & 39 & $P^2_2$ & 38 & $P^3_2$ & 49 & $Q^1_2$ & 40 & $Q^2_2$ & 45 & $Q^3_2$ & 35 \\ \hline
  $P^1_3$ & 30 & $P^2_3$ & 32 & $P^3_3$ & 36 & $Q^1_3$ & 35 & $Q^2_3$ & 35 & $Q^3_3$ & 45 \\
  \hline
\end{tabular}
\caption{parameters of the inverse demand functions considered in the example.} \label{Table_IDF}
\end{center}
\end{small}
\end{table}


\subsection{General assumptions}

We assume a central regulatory authority who has the exclusive right to sell pipeline capacity licenses
for market participants, who, according to their individual positions and demand functions, have different evaluations for particular routes in the network. Participants do have a strategy space -- they decide which bids they would like to submit, thus they can be considered as players of the game. For the clarification of the terminology, we will use 'participants' and 'players' as synonyms in the rest of the paper. In this paper we will assume that the exclusive participants of the capacity allocation game are consumers of the model, since, as mentioned earlier, their optimal bidding strategy may be plausibly derived from the modelling assumptions via simple computations in both of the mechanisms analyzed.

\subsection{Bidding and capacity allocation in the ascending clock auction (ACA) framework}
\label{ACA rules}

In the following we summarize the assumptions by which the bidding behavior in our model is described, and evaluate the ACC auction based on the principles laid down in \cite{EU2017-CommRegulation}.
The ACA auction process is carried out simultaneously for each line.
Let us summarize the critical points which are defined in the regulation, regarding our auction model.
\begin{itemize}
\item \emph{Ascending clock auctions shall enable network users to place volume bids against escalating prices announced in consecutive bidding rounds, starting at the reserve price $P_0$.}
\item \emph{The volume bid per network user at a specific price shall be equal to or less than the volume bid placed by this network user in the previous round.}
\item \emph{If the aggregate demand across all network users is less than or equal to the capacity offered at the end of the first bidding round, the auction shall close.}
\item \emph{If the aggregate demand across all network users is greater than the capacity offered at the end of the first bidding round or a subsequent bidding round, a further bidding round shall be opened with a price equal to the price in the previous bidding round, plus the large price step.}
\item \emph{If a first-time undersell occurs, a price reduction shall take place and a further bidding round shall be opened. The further bidding round will have a price equal to the price applicable in the bidding round preceding the first-time undersell, plus the small price step. Further bidding rounds with increments of the small price step shall then be opened until the aggregate demand across all network users is less than or equal to the capacity offered, at which point the auction shall close.}
\end{itemize}

In the above mechanism the clear aim of the large and small price steps is to reduce the number of bidding rounds (the auction switches to small price steps before the undersell). For the aim of simplicity, in our simulations we use only one step size, which is small enough to capture the details of the change of individual evaluations as the price of a certain line increases: It is easy to see that as every source cost, transfer cost, and demand function parameters are defined as integers, a step size smaller than the unit will not make any difference to the bidders in their evaluations (if the price started from an integer value). Thus in our simulations, we will assume that the price of every capacity product is increased by 1 in every step, if no undersell occurs.

In our simulation we assume that in the beginning, no capacities are allocated, every transfer capacity
is subject to the auction. This means that we will have 12 products: capacities corresponding to positive and negative directions for each edge.
According to the principles described in \cite{EU2017-CommRegulation}, the initial price of capacity products are set to cover only the expanses of the TSO. As the proposed model considers transfer prices separately (not included in capacity prices), this implies that we assume that the initial price of the capacity products is 0.

According to the considerations discussed before in subsection \ref{Motivation}, we simulate 3 rounds of ACA auctions, each with the starting price of one unit for each active product.

To clarify the terminology, each auction \emph{round} begins with the declaration of capacities under auction (and the initial prices of them, which is assumed to be 0), and after the submission of bids to every active product, the prices of overbidded products increase in every \emph{step} of the actual auction by 1 unit. If the total amount of submitted bids for a certain capacity are less than the volume of the particular capacity, the product becomes passive, and capacities are allocated according to the last submitted bids. The auction round ends, if all products become passive.

\subsubsection{Bid format in the ACA auction}

The bid format in the ACA is simple. In each bidding round, we have active products, for which bids may be still placed and closed products, for which the capacity is over. In the initial round all products are active.
In our case the products are the + and - directional transfer capacities of the pipelines, namely
$$
[q_1^+~~q_2^+~~q_3^+~~q_4^+~~q_5^+~~q_6^+~~q_1^-~~q_2^-~~q_3^-~~q_4^-~~q_5^-~~q_6^-~~]
$$

Each player must define the bid quantities he/she places on the active products in each round, in other words the overall bid of a player is defined by a vector.

In the next subsection we describe how we model the optimal bidding strategy of participants in these framework.

\subsubsection{Modelling of optimal bidding in the ACA framework}
\label{optimal_bidding_ACA}

\paragraph{Calculations during the steps of an ACA} Let us first emphasize that in the current model we assume that players have perfect information about source and transfer costs of the network (and of course about the actual prices of capacities in the auction), but they have no information about other player's utility functions. In other words, in the current simulation framework player do not make any speculations of other player's bids to optimize their own bidding strategy, they just consider the network parameters, the actual capacity prices in the particular step of the actual auction, and their own inverse demand function to determine their bids for the actual bidding step.

In the proposed modelling formalism we distinguish between \emph{already allocated capacities}, and \emph{potential capacities}. In a general case (e.g. if we are not considering the first bidding round), players may hold already allocated capacities. To be more precise, we assume that after the yearly auction finishes, each player assigns flows to the capacities, he/she obtained at the end of the auction to maximize its utility via fulfilling the demand in the respective consumer node. However, after the determination of these flows, some capacities may remain unused -- these are considered as free-to-use \emph{already allocated capacities} in the following (since the payment for them has already been completed). A simple example for such a scenario may be considered, if a player bids for several components (e.g. 2 line capacities) of a route, receives one of them in the early steps of the auction, but during the following steps the price of the other one increases so much that it does not make sense for the player to maintain its bid anymore. This way the player will be probably not able to assign any flow to this single capacity, thus he/she will have unused capacity at the end of the first round, which will be considered as \emph{already allocated capacity} in the second bidding round. Capacities still under auction are considered differently, since the payment for them has not been completed yet (using them implies additional cost, since they must be acquired first at their actual price).

To exactly determine the actual bids of a player in a given auction step, we use the principle of \emph{optimal potential flows}. This approach means the following. In each step of any auction round (1 2 or 3), players determine their optimal flows, which maximizes their resulting utility $U$, assuming that they will receive the capacities on which they place bids. The resulting utility may be calculated as the utility of consumption at the consumer node $U_C$, minus the cost of transfers ($C_T$) and inlets $(C_I)$, minus the payment for the capacity rights ($C_C$). In this calculation the player takes into account that flows planned on already allocated capacities do not imply furthers costs in addition to the transfer cost, in contrast flows planned on potential capacities imply additional cost, namely the actual capacity price.

In our model, players submit bids according to these optimal actual flows, namely we assume that they submit a bid vector, which is able to ensure the flows on potential capacities calculated in the optimal actual flows problem.

The formalism of the approach is the following. Let us consider a linear programming problem, as

\begin{align}
\label{eqs_opt_flows}
\max~U(x)~~\text{where}~~
x=\left(
    \begin{array}{c}
      f_{aac} \\
      f_{cau} \\
      I \\
      Y \\
    \end{array}
  \right)
  ~~s.t~~
  \begin{array}{c}
   0~\leq~f_{aac}~\leq~AAC \\
   0~\leq~f_{cua}~\leq~CUA \\
   0~\leq~f_{cua}~\leq~PB \\
   0~\leq~I\\ 
   0~\leq~ Y~\leq~\overline{Y}
  \end{array}
  ~~~
  A_{eq}x=0
\end{align}
In the above notation, $f_{aac}$ denotes the vector of flows on already allocated capacities, and may be decomposed as
\begin{align}
      f_{aac}=\left(
                \begin{array}{c}
                  f_{aac}^+ \\
                  f_{aac}^- \\
                \end{array}
              \right)
\end{align}
where $f_{aac}^+$ stands for the positive directional flows (according to the directions of edges) and
$f_{aac}^-$ denotes the negative directional flows. $f_{cau}$ denotes the flows on capacities under auction (potential capacities), and it has the same structure as $f_{aac}^-$.
$f_{aac}^+,~f_{aac}^-,~f_{cau}^+,~f_{cau}^-\in\mathcal{R}^m$, where $m$ is the number of edges (6 in the case of the proposed example).

$I\in \mathcal{R}^{n_s}$ denotes the vector of inlet values ($n_s$ stands for the number of source-nodes, and equals to 1 in the defined example), while $Y$ describes the consumption.
To each piecewise constant part of the inverse demand function of the respective player (these are depicted in Fig. \ref{NCA_example_DC_1}), we assign a variable $y_k$, so $Y \in \mathcal{R}^{n_p}$, where $n_p$ is the number of these piecewise constant parts ($n_p=3$ for all 3 players in our case).
As discussed previously, $\overline{I}$ is the maximum inlet, while

\begin{align}
\overline{Y}=\left(
               \begin{array}{c}
                 Q^i_1 \\
                 Q^i_2 \\
                 Q^i_3 \\
               \end{array}
             \right)~~~~~\text{so}~~~~
             Y=\left(
               \begin{array}{c}
                 y_1 \\
                 y_2 \\
                 y_3 \\
               \end{array}
             \right)
             \leq
             \left(
               \begin{array}{c}
                 Q^i_1 \\
                 Q^i_2 \\
                 Q^i_3 \\
               \end{array}
             \right)
\end{align}
where the $Q^i_1$ values depend on the actual player $i$.

The parameter vector $AAC$ holds the already allocated capacity values for each edge-direction pair.
$CUA$ stands for the vector of capacities currently under auction. $PB$ stands for the value of the previous bid vector submitted by the player. This constant is necessary, since the rules of the ACA discussed in subsection \ref{ACA rules} state that \emph{'volume bid per network user at a specific price shall be equal to or less than the volume bid placed by this network user in the previous round'}.

The equation $A_{eq}x=0$ formalizes the nodal balances. For each node, the inlets plus the inflows must be equal to the outflows plus the consumption. Using the variables in $x$, and the network topology, the $n$ equations corresponding to the rows of the $A_{eq}$ matrix may be easily derived ($n$ is the number of nodes).

Finally $U(x)$ stands for the resulting utility of the actual player, and may be decomposed as
$U=U_C-C_T-C_I-C_C$, where, as discussed previously, $U_C$ denotes the utility of consumption, $C_T$ and $C_I$ denote the cost of transfers and inlets, and $C_C$ denotes the payment for the capacity rights.
It is clear that $U_C$, $C_T$ and $C_I$ are linear functions of the variables of $x$ -- the coefficients of the linear functions may be derived from the $P^i_j$ values, the $c^t$ transfer costs (defined for the edges in Table \ref{Table_edge_par_EX1}), the inlet cost per unit (23 in our case). Finally, $C_C$ may be derived from the actual prices present in the auction step.

After each step, according to the potential allocations (if for any product, if there's no overbidding, the auction for that product is finished, it will be allocated at the actual price to the actual bidders), the values of already allocated capacities ($AAC$) and the capacities under auction (CUA) are updated.

\paragraph{Calculations between ACA rounds}

After a round of ACA auction has finished, each player calculates the optimal flows on the capacities actually allocated to him/her. This can be done easily by solving \ref{eqs_opt_flows}, under the assumption that there are no capacities under auction, only already allocated capacities. According to the results of this calculation, every player performs the two following operations:

\begin{enumerate}
  \item First, the player in question determines the quantity ensured by the calculated flows. As this quantity is ensured for him/her in the following, the player updates its inverse demand function for the remaining auction steps accordingly (demand is reduced by the already accessible quantity).
  \item Second, the player divides the capacities allocated to him/her into two groups: Capacities which are used by the flows are considered 'out of the game' in the following, thus they are allocated to the flows fixed after this round. Allocated capacities on the other hand which are not used in the flows are considered as already allocated capacities in the next auction round (payment for them has been already completed, thus they may be used in the design of potential optimal flows in the following).
\end{enumerate}

\subsubsection{Results of the ACA process for the simple example}

According to the principles discussed above, the three rounds of ACA auctions may be simulated for the proposed simple example. The details of the calculations can be found in Appendix A.

The resulting capacity allocation of the ACA process is described by the matrix detailed in E.q.
(\ref{example_CA_ACA}) (where rows correspond to players 1-3) and results in the resulting utility vector
$[U^1_{ACA}~~U^2_{ACA}~~U^3_{ACA}]=[ 720~~400~~ 1325]$. Thus the total utility of players ($U^T_{ACA}$) equals 2445 in this case.

\begin{small}
\begin{align}
\label{example_CA_ACA}
AC_{ACA}=\left(
  \begin{array}{cccccccccccc}
 0 &          0 &          0 &         10 &          0 &          0 &         80 &          0 &          0 &          0 &          0 &          0  \\
    0 &          0 &          0 &          0 &          0 &          0 &          0 &         60 &          0 &         10 &          0 &          0  \\
    0 &          0 &          0 &          0 &          0 &          0 &          0 &         15 &         70 &          0 &          0 &         15  \\
  \end{array}
\right)
\end{align}
\end{small}

Let us introduce some further characteristic values as bases for latter comparison. The total payment received by the auctioneer ($\sum C_C$) for the auctioned capacities is 700 units in this case.
We can furthermore characterize how much the available infrastructure is utilized. If we consider the total transfer capacity of the network versus the amount of capacity allocated via the process and versus the capacity actually used by the players at the end in their resulting optimal flows on their available capacities, we can calculate that the ratio of allocated network capacities ($r^{ANC}$) is 0.321 while the ratio of network capacities which are used in the end ($r^{UNC}$) is 0.2963 (let us note that one edge is typically used in one direction, but capacity products for both directions are present in the auction, thus total usage ratios over 0.5 are very unlikely).

In the following we introduce the convex combinatorial auction and demonstrate its functioning and results in the case of the introduced example.

\subsection{The convex combinatorial auction (CCA) for the allocation of network capacities}

As we will see, the CCA framework uses a route-centered formalism. Our first task in this approach is to define the routes of players, via which they are potentially able to transport the gas for themselves.
We consider the following routes for player 1 2 and 3 respectively:
\begin{table}[h!]
\begin{center}
\begin{tabular}{|c|c|}
  \hline
  $R^1_1$ & -1 \\ \hline
  $R^1_2$ & -2,~4 \\ \hline
  $R^1_3$ & -3,~5 \\ \hline
  $R^1_4$ & -2,~-6,~5 \\ \hline
  $R^1_5$ & -3,~6,~4 \\
  \hline
\end{tabular}
~~~
\begin{tabular}{|c|c|}
  \hline
  $R^2_1$ & -2 \\ \hline
  $R^2_2$ & -1,~-4 \\ \hline
  $R^2_3$ & -3,~5,-4 \\ \hline
  $R^2_4$ & -3,~6 \\
  \hline
\end{tabular}
~~~
\begin{tabular}{|c|c|}
  \hline
  $R^3_1$ & -3 \\ \hline
  $R^3_2$ & -1,~-5 \\ \hline
  $R^3_3$ & -2,~-6 \\ \hline
  $R^3_4$ & -2,~4,-5 \\
  \hline
\end{tabular}
\caption{Routes of players considered in the proposed example. Every route is a set of edges leading from the source to the consumer. The signs are positive if the direction of the edge coincides with the route.}
\end{center}
\end{table}

In this setup we suppose that players may submit bids for route-quantity pairs, according to the principle that in the outcome of the auction a convex combination of their submitted bids will be assigned to them.
This assumption allows bidding for alternative routes: If a consumer needs 1 unit of gas and there are 2 alternative sources in the network, corresponding to two different routes, he/she can submit two bids for the capacity licenses of the two distinct routes, both with the quantity of 1 unit. At the end of the auction a convex combination of the two bids will be assigned to him, which means that he/she will not get more network capacity towards the sources than 1 unit, but this maximally 1 unit may be composed of arbitrary proportion of the two routes. Of course this line of thought applies for a single source with multiple access routes as well.

\subsubsection{Bid format in the CCA auction}
\label{Bid format}

As mentioned earlier, in the CCA framework participants of the auction may submit bids for route-quantity pairs. Let us denote the $k$-th bid for route $j$ of player $i$ with $B^i_{j,k}=(q^i_{j,k},~p^i_{j,k})$ where $q^i_{j,k}$ is the quantity of the bid, and $p^i_{j,k}$ is the price offered for the route-quantity pair in question.

\subsubsection{Variables of the clearing problem}
$x^i_{j,k} \in [0,1]$ denotes the acceptance indicator of the bid $B^i_{j,k}$.

\subsubsection{Constraints}

\paragraph*{Network constraints:} To formulate the constraints which describe the limited capacity of pipelines, we need to decompose the routes considered in the auction to their components -- to edges which correspond to pipelines. Furthermore we take into account the possibility that counter-directed flows cancel each other.

Let us denote the set of (directed) edges in the network with $E$, while $e \in E$ denotes a single edge.
Each route $j$ (of player $i$) may be represented as an $R^i_j \subseteq E$ subset of edges, where each element is signed, according to whether the direction of the route coincides with the direction of the included edge or not.

Let us suppose furthermore that edge $e_m$ has different maximal capacity in the positive and negative direction (think of one-directional pipelines), denoted by $\bar{q}_m^+$ and $\bar{q}_m^-$ respectively.

In this case the maximal capacity constraints may be formulated as

\begin{align}
\sum_{i,j,k~~e_m \in R^i_j}s^i_{j,m}~x^i_{j,k}~q^i_{j,k} \leq \bar{q}_m^+~~~~~\sum_{i,j,k~~e_m \in R^i_j} - s^i_{j,m}~x^i_{j,k}~q^i_{j,k} \leq \bar{q}_m^- ~~~~~~~~ \forall m
\end{align}

where $s^i_{j,m}$ is an indicator variable, which equals to 1 if edge $e_m$ has positive sign in route $j$ of player $i$, and -1 otherwise.

Maximal output limitations of sources in the network may be derived very similarly by constraining the total outflow of the edges connected to the source in question.

\paragraph*{Convexity constraint:}

By definition, the auction assigns to each player a convex combination of his/her submitted bids.
This consideration is formalized as

\begin{align}
\sum_{j,k} x^i_{j,k} \leq 1 ~~~~~~~ \forall i
\end{align}


\subsubsection{The optimization problem of the CCA framework}

The objective of the optimization process is to maximize the nominal income from the bids, under the previously detailed constraints.

\begin{align}
\label{CCA_optimization_problem}
\begin{array}{c}
 \max_{x}~~~ x^i_{j,k}p^i_{j,k}~~~~~\text{s.t.}\\
 \\
 \sum_{i,j,k~~e_m \in R^i_j}s^i_{j,m}~x^i_{j,k}~q^i_{j,k} \leq \bar{q}_m^+~~~~\forall m \\
 \\
\sum_{i,j,k~~e_m \in R^i_j} - s^i_{j,m}~x^i_{j,k}~q^i_{j,k} \leq \bar{q}_m^- ~~~~\forall m \\
\\
\sum_{j,k} x^i_{j,k} \leq 1 ~~~~~~~ \forall i \\
 \end{array}
\end{align}

The above problem falls into the class of linear programming problems. Let us recall that regarding the ACA framework, we only used linear programming to model optimal bidding behavior, but the allocation itself in that case has been performed by a logical algorithm described in subsection \ref{ACA rules}.
In contrast, in the case of CCA, the allocation process itself relies on solving a linear programming problem. Let us point out here however that in other auction framework related to energy economics as
electricity auctions linear programming, and even more computationally demanding programming problems (as integer and quadratic programming) are routinely used in practice (see e.g. \cite{madani2017revisiting}).

\subsection{Payments}
\label{subsection_VCG}

 After the optimization process has been completed and the bid acceptance ratios have been determined, the payments of the players have to be completed. To determine payments in the proposed framework, we use the
 Vickrey-Clarke-Groves (VCG) mechanism \cite{vickrey1961counterspeculation,groves1973incentives,scherr1975pricing}, which charges each individual the harm they cause to other bidders
 with their participation. The VCG mechanism gives bidders an incentive to bid their true valuations, by ensuring that the optimal strategy for each bidder is to bid their true valuations of the items. It is a generalization of a Vickrey auction \cite{vickrey1961counterspeculation} for multiple items.
 In the following, we apply the proposed CCA framework for the simple example, and also detail how the VCG-payments are calculated.

\subsection{Results of the CCA process for the simple example}

The first step is to derive the bids of the CCA approach. This may be done via simple calculations, which are detailed in Appendix B. Considering the bids detailed in in Appendix B and calculating the optimum of the auction problem summarized in (\ref{CCA_optimization_problem}), we get the following results.

\begin{itemize}
  \item Regarding player 1, the nonzero acceptance indicators are $x^1_{1,1}=0.5$ and $x^1_{1,2}=0.5$ resulting in the total capacity of 70 units on route $R^1_1$ of player 1.
  \item Regarding player 2, the nonzero acceptance indicators are $x^2_{1,1}=0.6667$ and
  $x^2_{1,2}=0.3333 $, resulting in the capacity of 55 on route $R^2_1$ of player 2.
  \item Regarding player 3, the relevant indicators are $x^3_{1,2}=0.8235$ and $x^3_{3,2}=0.1765$, resulting in the capacities of 70 and 15 on routes $R^3_1$ and $R^3_3$.
\end{itemize}

From the resulting acceptance indicators, and from the bid and route data, we can determined the allocated edge capacities

\begin{small}
\begin{align}
\label{example_CA_CCA}
AC_{CCA}=\left(
  \begin{array}{cccccccccccc}
 0 &          0 &          0 &          0 &          0 &          0 &         70 &          0 &          0 &          0 &          0 &          0  \\
    0 &          0 &          0 &          0 &          0 &          0 &          0 &         55 &          0 &          0 &          0 &          0  \\
    0 &          0 &          0 &          0 &          0 &          0 &          0 &         15 &         70 &          0 &          0 &         15  \\
  \end{array}
\right)
\end{align}
\end{small}
If we compare the resulting capacity allocation of the CCA in eq. (\ref{example_CA_CCA}) to the resulting allocation of the ACA in eq. (\ref{example_CA_ACA}), we can already see that the two methods result in different allocations.

\subsubsection{Determination of payments}
As detailed in subsection \ref{subsection_VCG}, we use the VCG mechanism for the determination of capacity payments. The payment of any player is equal to the harm its participation in the auction implies for other players. The 'harm' in our case is measures in the cumulative nominal value of accepted bids for each player. Let us consider player 1.

In this case we have to first calculate the total nominal value of (at least partially) accepted bids for players 2 and 3. If we consider the bid acceptance indicators detailed above, and the CCA bid values detailed in Appendix B, we can see that the total nominal bid value of Player 2 ($TNBV^2$) may be calculated as
$$
TNBV^2=\frac{2}{3}600 + \frac{1}{3}915 = 705
$$
Similarly for Player 3,
$$
TNBV^2=0.8235 \cdot 1475 + 0.1765 \cdot 1305 = 1445~~~~~~,
$$
thus the total nominal bid value of Players 2 and 3 is 2150 in this case.

If we would like to know how much harm is implied by Player 1 to Players 2 and 3 by its participation, we have to simply re-run the CCA allocation process, assuming no bids for player 1.
In this case we get the following results.

\begin{small}
\begin{align}
& x^2_{1,2}=0.8235 ~~~~~ x^2_{2,2}=0.1765 \\
& x^3_{1,2}=0.8235 ~~~~~ x^3_{2,2}=0.1765~~~~~,
\end{align}
\end{small}
implying the quantity of 70 on routes $R^2_1$ and $R^3_1$ and the quantity of 15 on routes $R^2_2$ and $R^3_2$. The reason for this symmetry is that the if we consider the sum of the quantities of the first two steps in the demand curves, we get 85 for both players 2 and 3 (40+45 vs 50 + 35).

Let us now calculate the modified total nominal bid values (denoted by $TNBV_m$)
\begin{small}
\begin{align}
&TNBV^2_m=0.8235 \cdot 915 + 0.1765 \cdot 490 = 839.98 \\
&TNBV^3_m=0.8235 \cdot 1475 + 0.1765 \cdot 1262.5 = 1437.5~~~~~,
\end{align}
\end{small}
thus the total nominal bid value of Players 2 and 3 is appr. 2277.5 in  this case.

According to the VCG principles described in subsection \ref{subsection_VCG}, the capacity payment of Player 1 is 2277.5-2150=127.5.

This process may be straightforwardly repeated for players 2 and 3 to determine the capacity payments.
In this case we get the result
$$
[C^1_C~~~C^2_C~~~C^3_C]=[127.5~~~80~~~115]
$$

\subsubsection{Results of the CCA process for the simple example}

After the allocated capacities have been determined, the optimal flows on the available capacities may be calculated for each player. This can be done similarly to the case described in Appendix A in the ACA case, but in this case the optimal flows will correspond exactly to the (possibly partially) accepted bid-quantity pairs on the respective routes. Knowing the inlet costs, the transfer costs, the capacity costs and the resulting consumption values, utility calculation may be carried out the same way as in the case of the ACA, described in Appendix A.
\begin{small}
\begin{align}
&U^1_{CCA}=U^1_C - C^1_T - C^1_I -C^1_C=3130    - 630  - 1610  - 127.5 =762.5  \nonumber \\
&U^2_{CCA}=U^2_C - C^2_T - C^2_I -C^2_C=2410    - 440  - 1265  - 80 =625 \nonumber \\
&U^3_{CCA}=U^3_C - C^3_T - C^3_I -C^3_C=4365    - 965  - 1955  - 115 =1330
\end{align}
\end{small}
In this case the total utility of players is $U^T_{CCA}=2717.5$ in contrast to the ACA case where, as we have seen, $U^T_{ACA}=2320$.

The total payments for the capacity rights in this case is 322.5 in contrast to the value of 825 in the ACA case, while the ratios of allocated network capacities and used network capacities ($r^{ANC}_{CCA}$ and $r^{UNC}_{CCA}$) are both 0.2778, in contrast to the values $r^{ANC}_{ACA}=0.321 $ and $r^{UNC}_{ACA}=0.2963$ calculated in the ACA case (there are no unused capacities in the CCA case).

Naturally, this example is not a sufficient basis for reaching general conclusions about the properties of the two methods, but is has been useful to demonstrate our models in detail. in the next section we
provide a simulation-based computational analysis for the statistical comparison of the two methods.

%

\section{Results}
\label{sec_results}

We used a computational approach to compare the performance of the ACA end CCA methods.
We generated random setups and simulated the capacity allocation processes to determine the resulting capacity allocations and payments of the two methods.

Each setup was generated as follows.
Input parameters were:
\begin{itemize}
  \item The number of vertices (nodes) $n_v$
  \item The number of edges $n_e$
  \item The number of sources $n_s$
  \item Upper and lower bounds for edge capacities $\overline{q}^{max}$ and $\overline{q}^{min}$
  \item Upper and lower bounds for transfer costs $C_T^{max}$ and $C_T^{min}$
  \item Upper and lower bounds for source costs $C_S^{max}$ and $C_S^{min}$
\end{itemize}

In the first step, graph of the network was generated. The first edge was placed randomly, the second was placed randomly among unconnected node-pairs and so on, until all $n_e$ edges have been placed (see Erd\H{o}s-R\'{e}nyi graphs \cite{erdHos1960evolution}). At the end, connectedness and planarity of the resulting graph was checked, and if any property did fail, the process was started over. Once the graph proved to be appropriate, $n_s$ nodes were picked at random from the set of nodes, and they were defined as source nodes (the rest are considered as consumer nodes in the following).

In the second step, the parameters of edges were determined. Transfer costs for edges were assumed to be identical in any direction, thus $n_e$ random values from a uniform distribution between $\overline{q}^{min}$ and $\overline{q}^{max}$ were picked, and rounded to the closest integer value to determine maximal edge capacities, and similarly, random values from a uniform distribution between $C_T^{min}$ and $C_T^{max}$ were picked, and rounded to the closest integer value to determine edge transfer costs.

Following this step, the (maximum) 10 cheapest source-consumer routes were determined for every consumer, considering source and transfer costs as well. The minimal and maximal values of these routes ($C_{Route}^{min}$ and $C_{Route}^{max}$) were calculated from the results.

In the following, supply and demand parameters were set as follows. $n_s$ random values from a uniform distribution between $C_S^{min}$ and $C_S^{max}$ were picked, and rounded to the closest integer value to determine the source costs.

Inverse demand functions were determined as follows. We assumed the three-step piecewise constant form as
in subsection \ref{consumer_demand}, where the quantity of each step was determined by picking a random integer value from the interval [10,~50]. The price of each step was determined by picking a random integer value from the interval [$C_{Route}^{min}+C_S^{min}$,~$1.4 \cdot (C_{Route}^{max}+C_S^{max})$].

We considered various network sizes each with different node, edge and source node numbers ($n_v,~n_e,~n_s$), but the other parameters were fixed as summarized in Table \ref{sim_pars}. For each network size, 1000 setups were generated, on which the ACA and CCA methods have been evaluated.

\begin{table}[h!]
\begin{center}
\begin{tabular}{|c|c|c|c|}
  \hline
  par. & value & par. & value \\ \hline
  $\overline{q}^{min}$ & 10 & $\overline{q}^{max}$ & 90 \\
  $C_T^{min}$ & 3 & $C_T^{max}$ & 11 \\
  $C_S^{min}$ & 20 & $C_S^{max}$ & 30 \\
  \hline
\end{tabular}
\caption{Invariant parameters used in the simulations \label{sim_pars}}
\end{center}
\end{table}

\subsection{Results on 6-node networks}
In this case $n_v=6,~~n_e=8,~~n_s=1$ were assumed. This resulted in an average route length of
1.56 between consumers and sources in these networks.

\begin{figure}[h!]
  \centering
  \includegraphics[width=6cm]{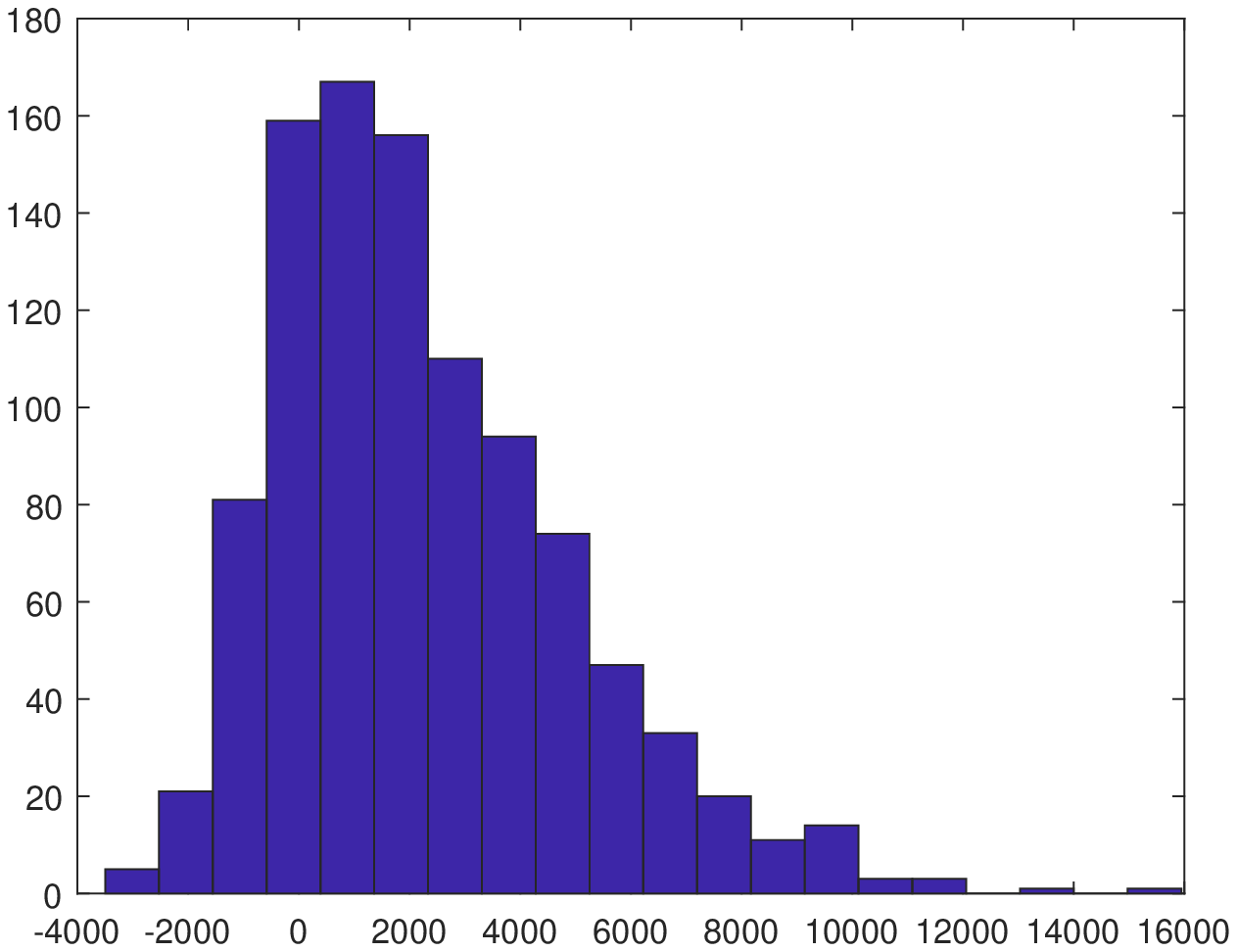}~
  \includegraphics[width=6cm]{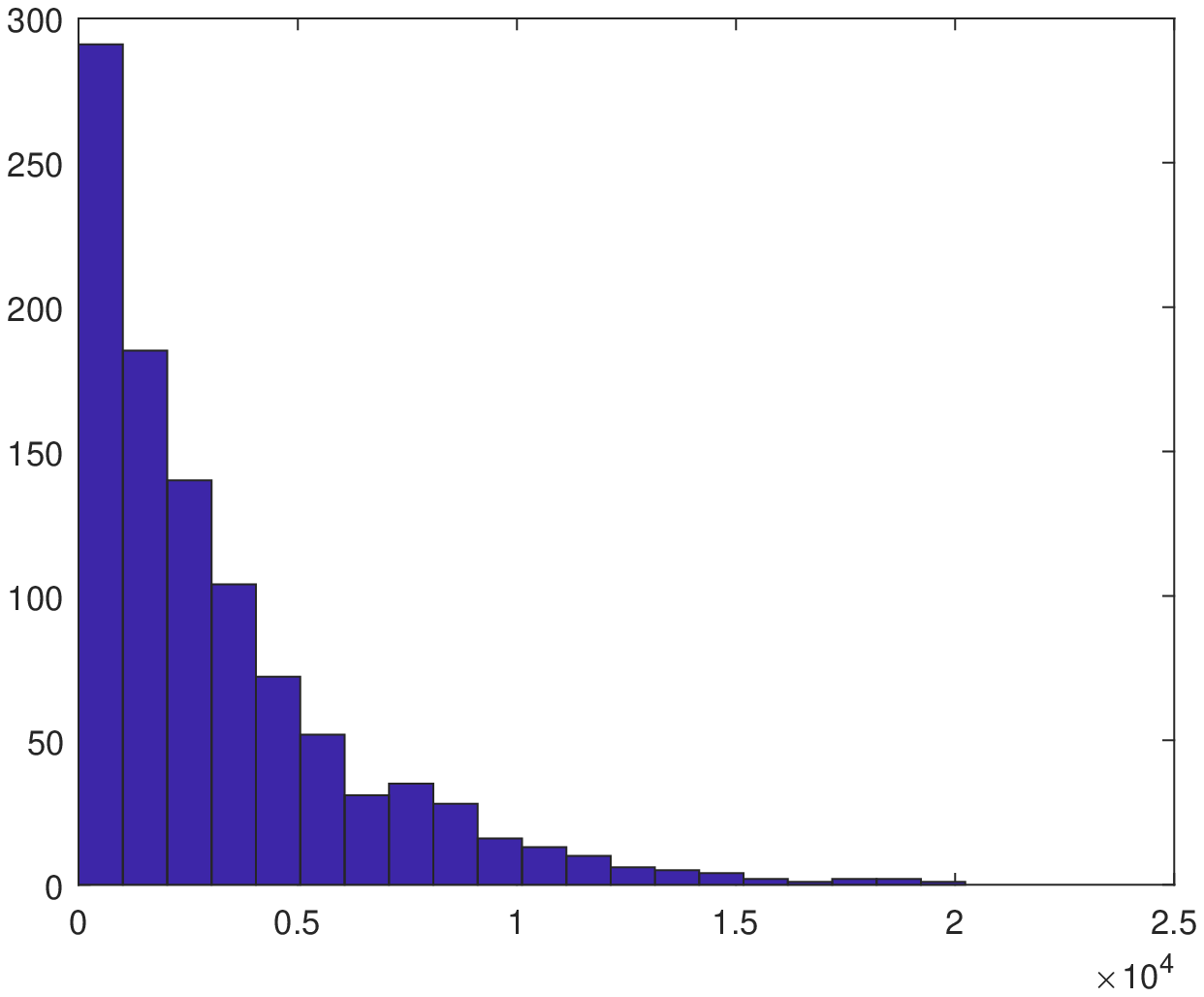}
  \includegraphics[width=6cm]{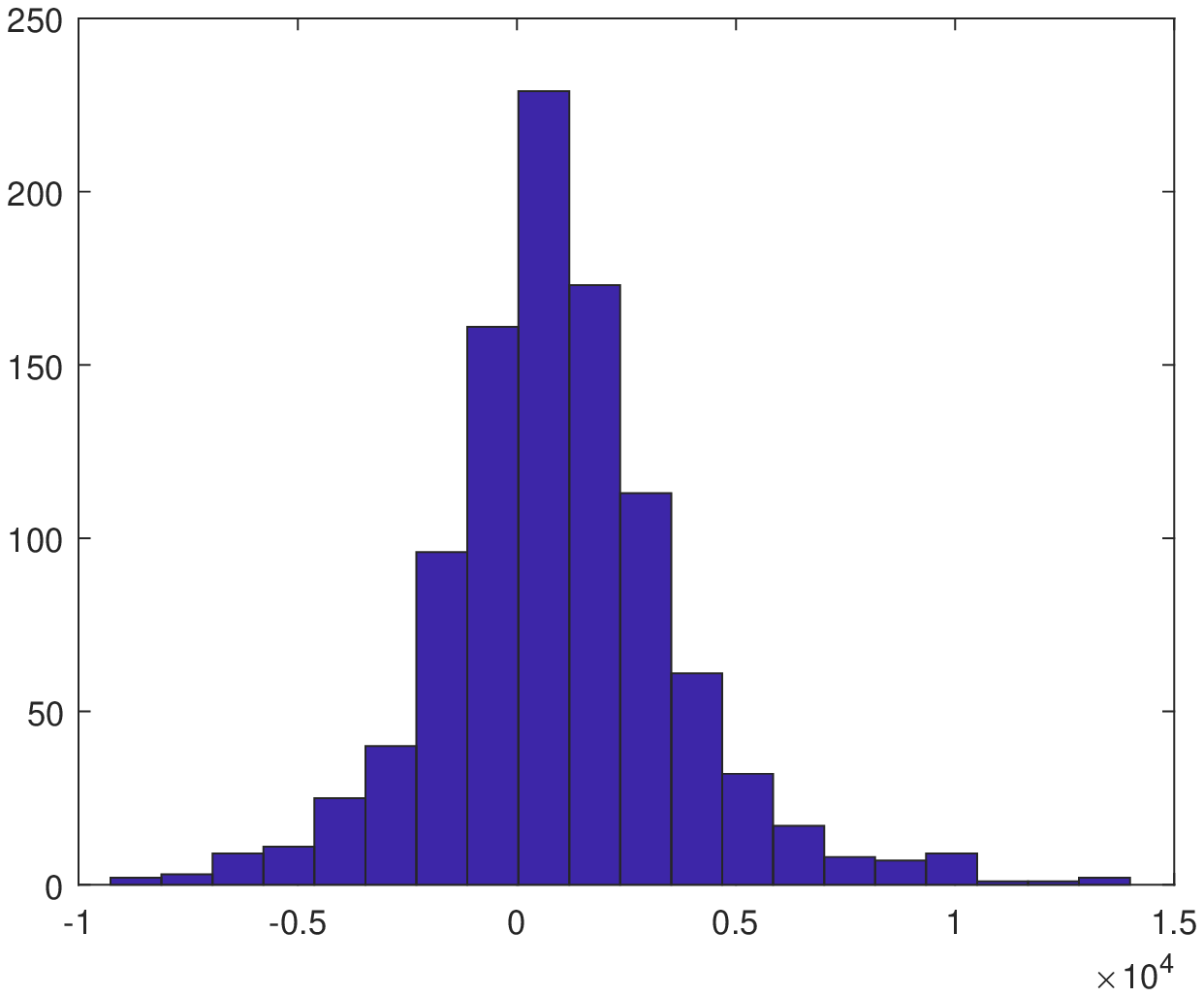}
\caption{Distributions of $U^T_{ACA}$, $U^T_{CCA}$ and of $U^T_{CCA}-U^T_{ACA}$ over the 1000 simulations done with 6-node networks.}\label{U_hist_6_node}
\end{figure}

Regarding the total resulting utility values and their difference ($U^T_{CCA}-U^T_{ACA}$), Fig. \ref{U_hist_6_node} depicts the distributions over
the 1000 simulated cases.

The first thing we may notice is that in the ACA case there are scenarios, where $U^T_{ACA}$ is negative.
The reason for this is that in the ACA framework, it is possible that in the process of capacity allocation such capacities will be allocated to players, which will be useless for them as later they are determining their optimal flows on the capacities assigned to them. We have seen this phenomenon in the simple example, where at the end of the ACA process a capacity of 10 units on edge 3 has been assigned (see the matrix in eq. (\ref{example_CA_ACA})). Capacity payments for these unused capacities which do not form a full route at the end of the process imply negative utility components for these players. If these negative components outweigh the positive ones in the context of all players, $U^T_{ACA}$ may be negative as well. This happens in the 19 \% of the cases when we apply ACA.

Average values of $U^T_{ACA}$ and $U^T_{CCA}$ are 2267.5 and 3213.1 in this case, while their standard deviations are  2660.6  and 3224.7 respectively.
As we see in Fig. \ref{U_hist_6_node}, the CCA method does not perform better in every case.
In the 34\% of cases, the ACA allocation method results in higher $U^T$ values.

Regarding the income of the auctioneer, the average total capacity payments are
5424.3 and 4673 in the ACA and CCA cases respectively.

In addition, let us compare the ratios of the allocated and used network capacities ($r^{ANC}$ and $r^{UNC}$).
$$
r^{ANC}_{ACA}=0.4601~~~~r^{ANC}_{CCA}=0.2489~~~~r^{UNC}_{ACA}=0.2286~~~~r^{UNC}_{CCA}=0.2489
$$

Regarding the 'fairness' of the auction method, several approaches and measures can be used.
In this work we restrain ourselves to a very simple indicator regarding this aspect.
For each simulation, we can calculate the difference of the maximal and minimal resulting utilities
among players, and average this value over the simulations. These 'unfairness' indicator ($UF$)
results in the following values in the case of 6-node networks.
$$
UF_{ACA}= 1772.5~~~~UF_{CCA}=1390.8
$$

Finally, let us note that the ACA method reached its final values in the 77.5\% of the simulated cases (in these cases no bids were submitted for the 3rd round).

\subsection{Results on 9-node networks}
In this case $n_v=9,~~n_e=12,~~n_s=2$ were assumed, resulting in an average route length of 1.8739 between sources and consumers.

\begin{figure}[h!]
  \centering
  \includegraphics[width=6cm]{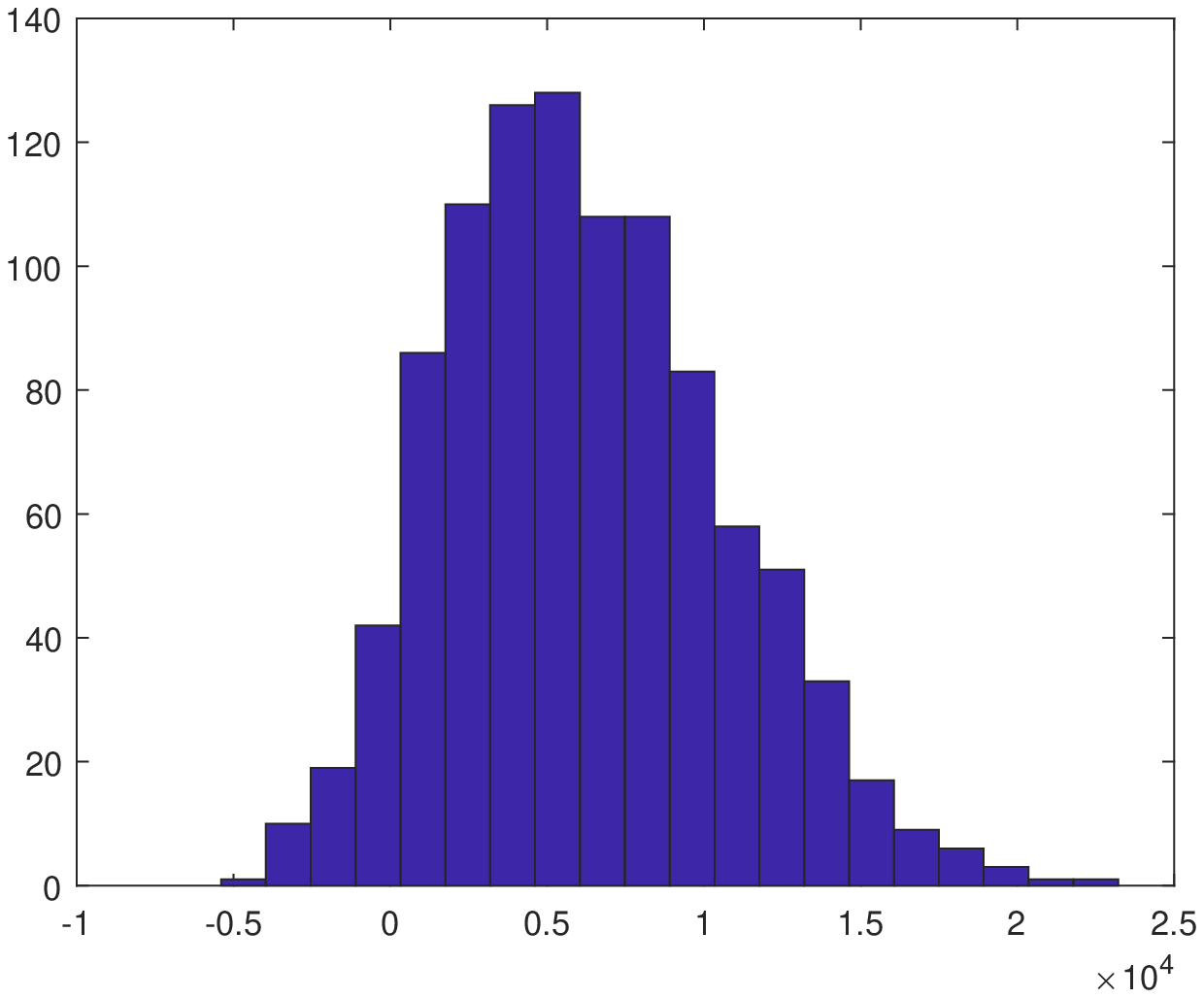}~
  \includegraphics[width=6cm]{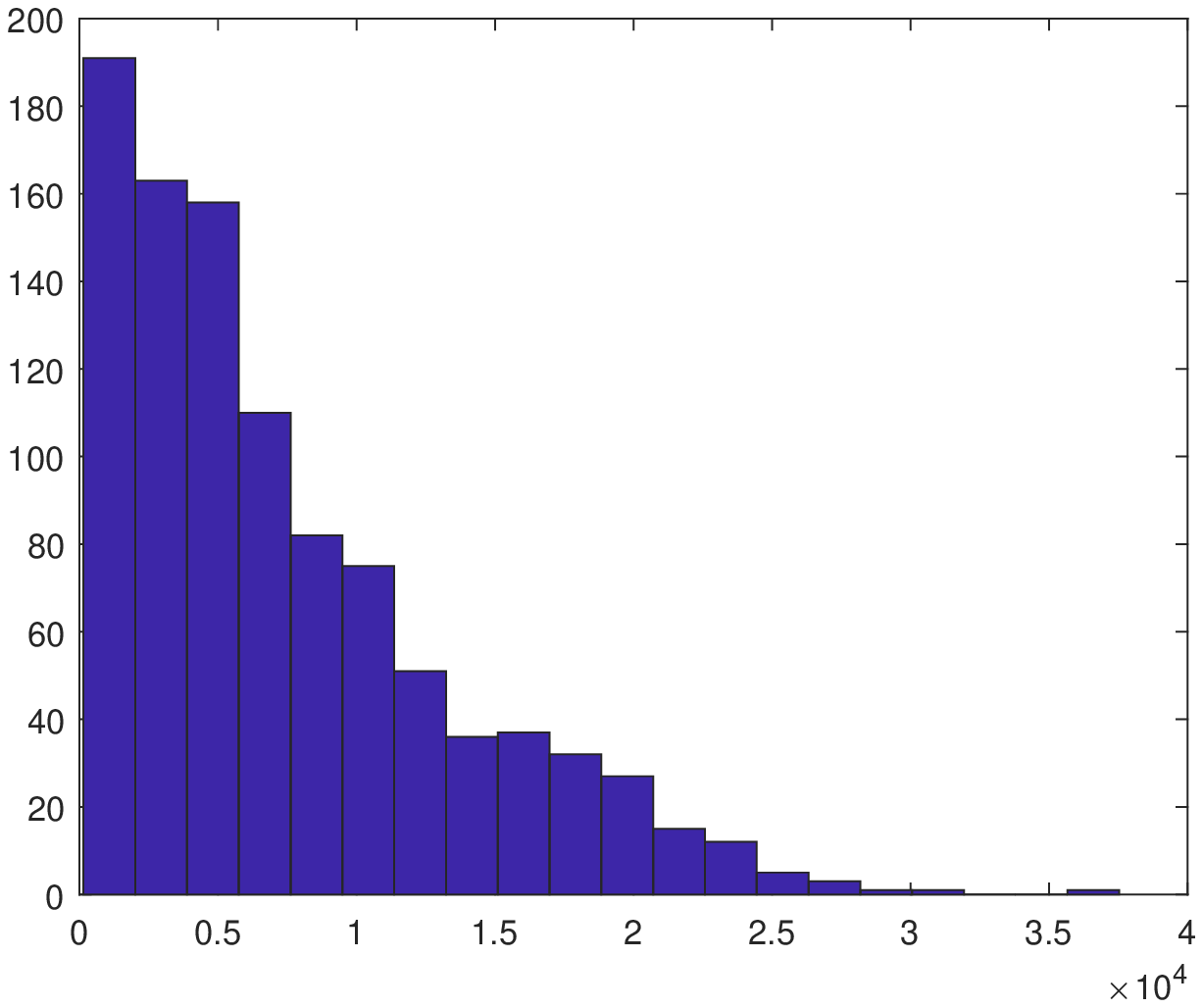}
  \includegraphics[width=6cm]{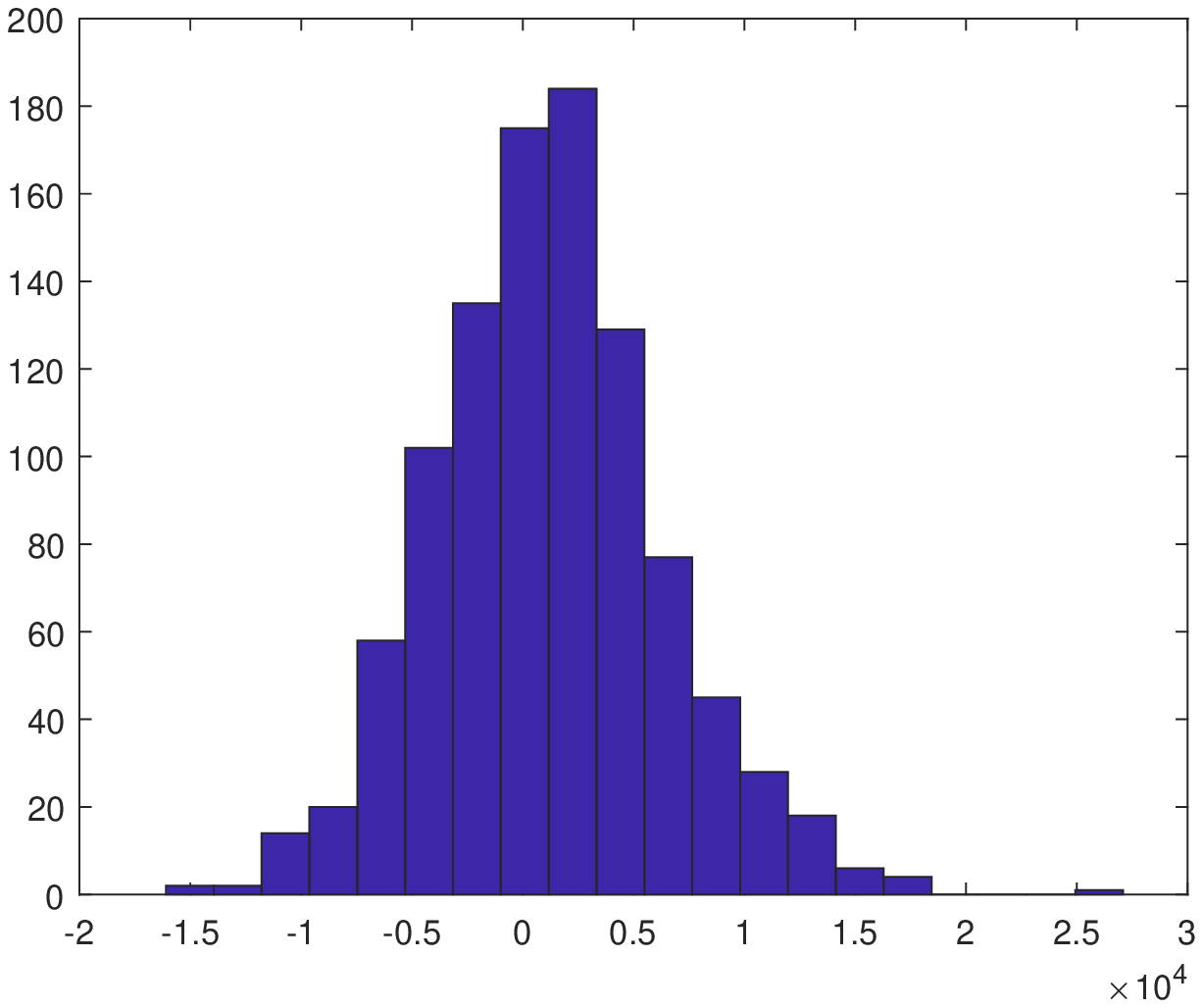}
\caption{Distributions of $U^T_{ACA}$, $U^T_{CCA}$ and their difference over the 1000 simulations done with 9-node networks.}\label{U_hist_9_node}
\end{figure}

Regarding the total resulting utility values, Fig. \ref{U_hist_9_node} depicts the distributions over
the 1000 simulated cases.

Average values of $U^T_{ACA}$ and $U^T_{CCA}$ are 6231.9 and  7407.3 in this case, while their standard deviations are 4440 and 6105.7 respectively.
The ACA method resulted in negative total utilities in the 5.9 \% of the simulated cases.
Regarding this network size, the ACA allocation method results in higher $U^T$ values in the 40.6\% of simulated cases.

Regarding the income of the auctioneer, the average total capacity payments are
8240 and 7221.1 in the ACA and CCA cases respectively.

Let us note that as the network size increases, the average length, thus the average total cost of routes is also increased. Since according to the simulation assumptions, the inverse demand functions are determined based partially on the route costs, they are also affected. These considerations explain the increasing trend in the utility values and capacity payments.

The ratios of the allocated and used network capacities ($r^{ANC}$ and $r^{UNC}$) are as follows.
$$
r^{ANC}_{ACA}=0.4678~~~~r^{ANC}_{CCA}=0.28~~~~r^{UNC}_{ACA}=0.2601~~~~r^{UNC}_{CCA}=0.28
$$

The 'unfairness' indicators ($UF$) are as
$$
UF_{ACA}= 3339.8~~~~UF_{CCA}=2496.0~~~~.
$$

The ACA method reached its final values in the 47.3\% of the simulated cases (in these cases no bids were submitted for the 3rd round).

\subsection{Results on 15-node networks}
In this case $n_v=15,~~n_e=20,~~n_s=3$ were assumed, resulting in an average route length of 2.503 between sources and consumers.

\begin{figure}[h!]
  \centering
  \includegraphics[width=6cm]{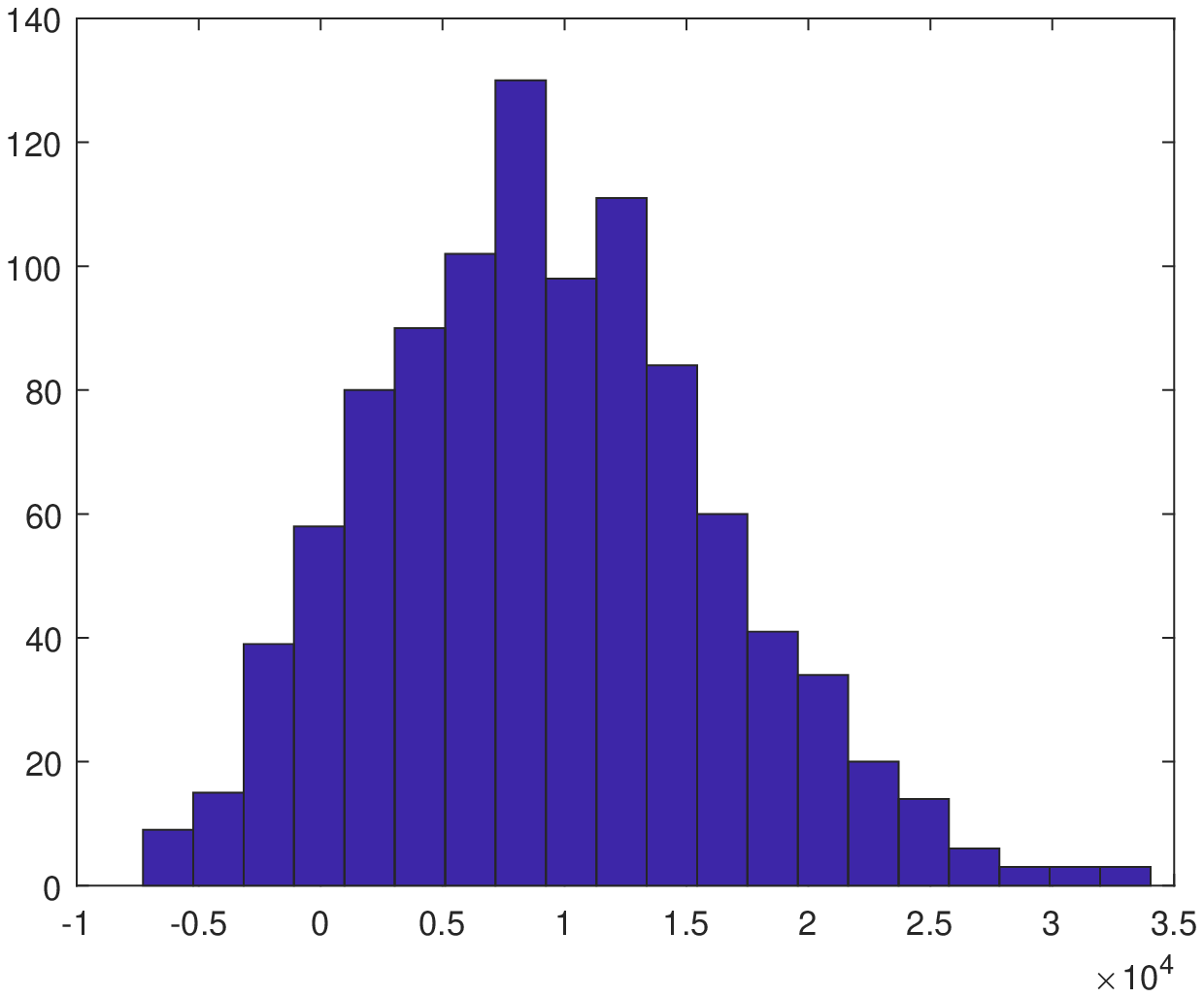}~
  \includegraphics[width=6cm]{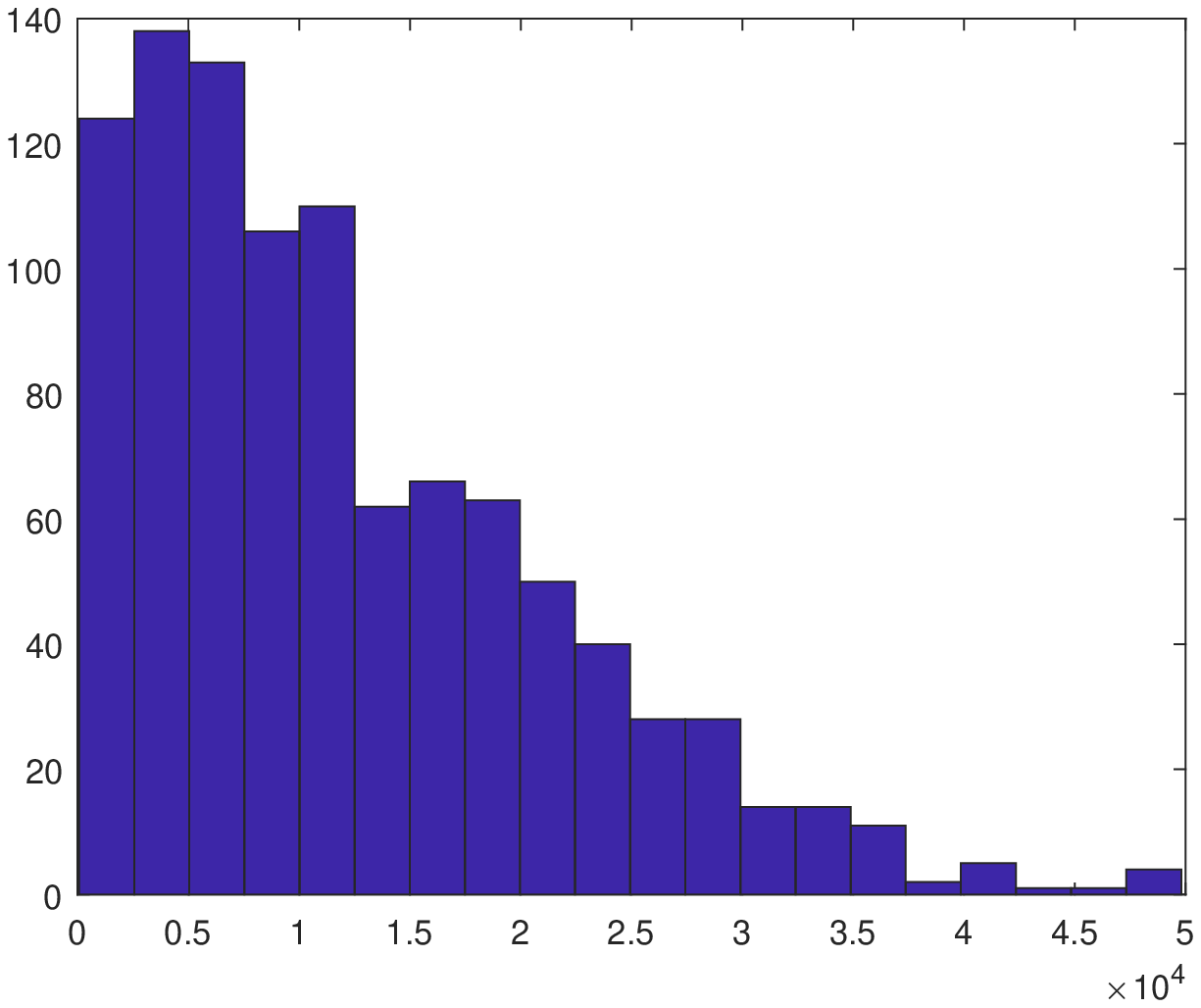}
  \includegraphics[width=6cm]{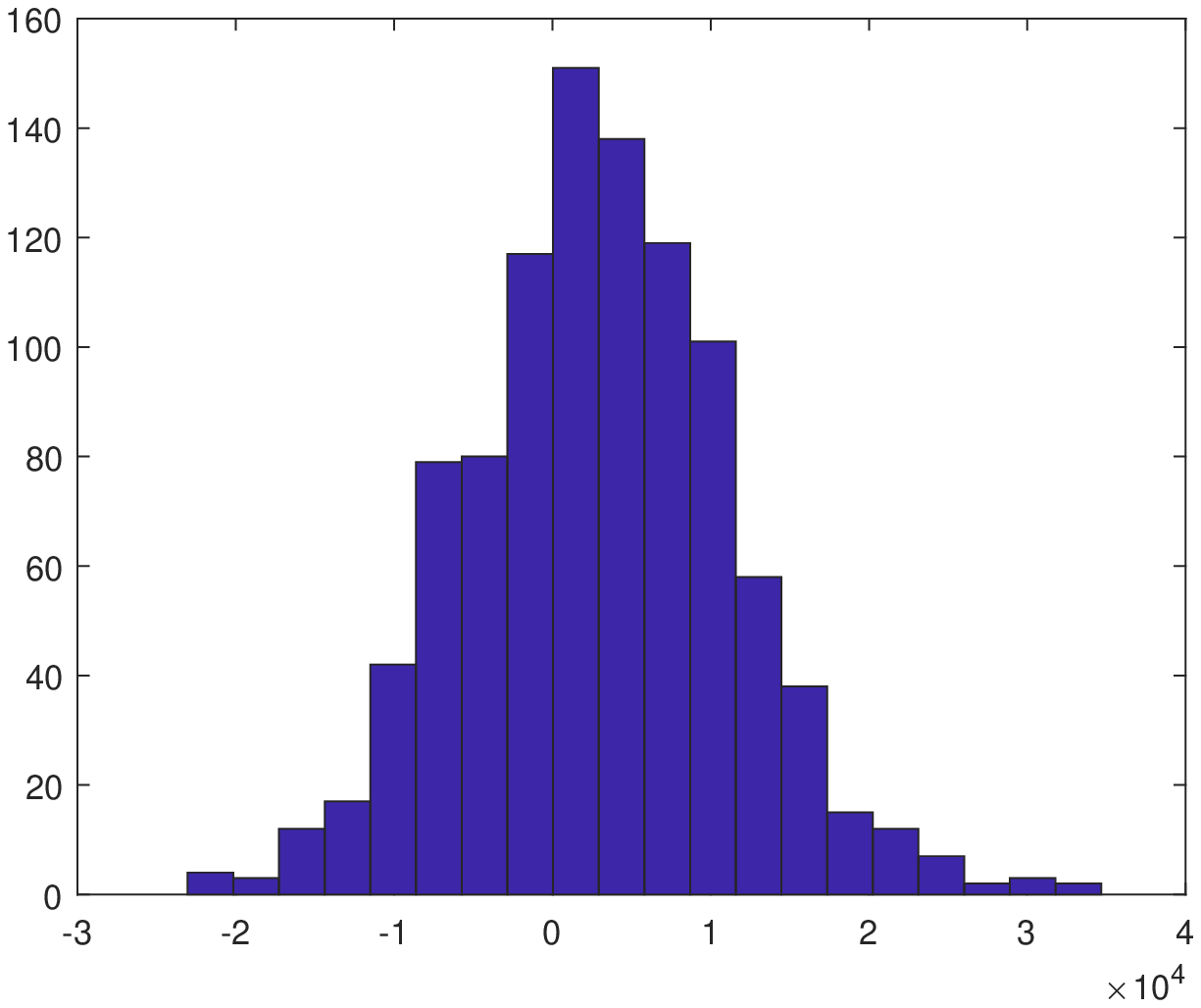}
\caption{Distributions of $U^T_{ACA}$, $U^T_{CCA}$  and their difference over the 1000 simulations done with 15-node networks.}\label{U_hist_15_node}
\end{figure}

Regarding the total resulting utility values, Fig. \ref{U_hist_15_node} depicts the distributions.

Average values of $U^T_{ACA}$ and $U^T_{CCA}$ are 9272 and 12223 in this case, while their standard deviations are  7077 and 9353 respectively.
The ACA method resulted in negative total utilities in the 9.6 \% of the simulated cases.
Regarding this network size, the ACA allocation method results in higher $U^T$ values in the 35.3\% of simulated cases.

Regarding the income of the auctioneer, the average total capacity payments are
 16555  and 14499 in the ACA and CCA cases respectively.

The ratios of the allocated and used network capacities ($r^{ANC}$ and $r^{UNC}$) are as follows.
$$
r^{ANC}_{ACA}=0.4928~~~~r^{ANC}_{CCA}=0.2758~~~~r^{UNC}_{ACA}=0.2416~~~~r^{UNC}_{CCA}=0.2758
$$

The 'unfairness' indicators ($UF$) are as
$$
UF_{ACA}= 5124.4~~~~UF_{CCA}=3490.8~~~~.
$$

The ACA method reached its final values in the 24.2\% of the simulated cases (in these cases no bids were submitted for the 3rd round).

\subsection{Results on 20-node networks}
In this case $n_v=20,~~n_e=30,~~n_s=4$ were assumed, which resulted in an average route length of
 2.3514 between sources and consumers.

\begin{figure}[h!]
  \centering
  \includegraphics[width=6cm]{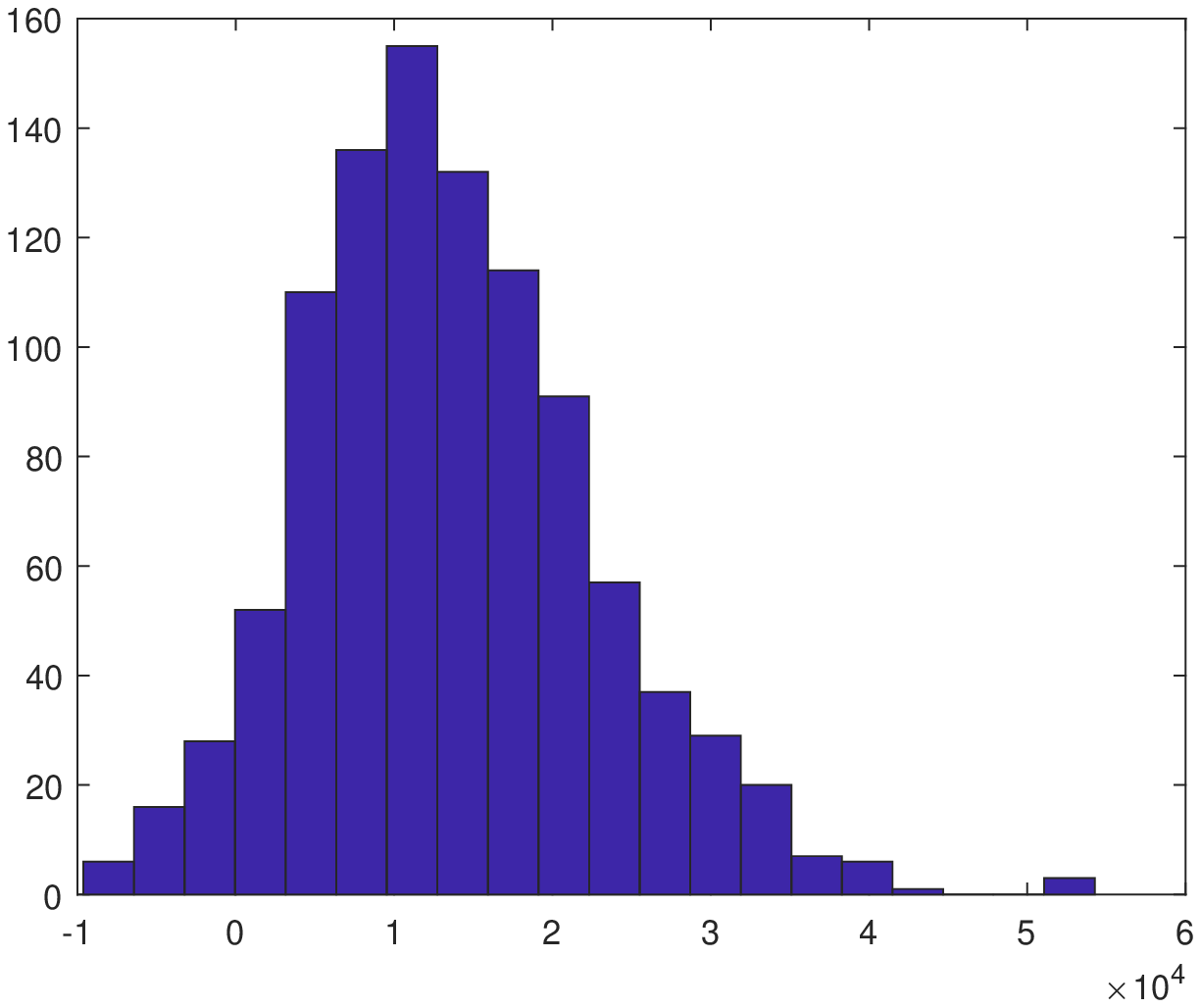}~
  \includegraphics[width=6cm]{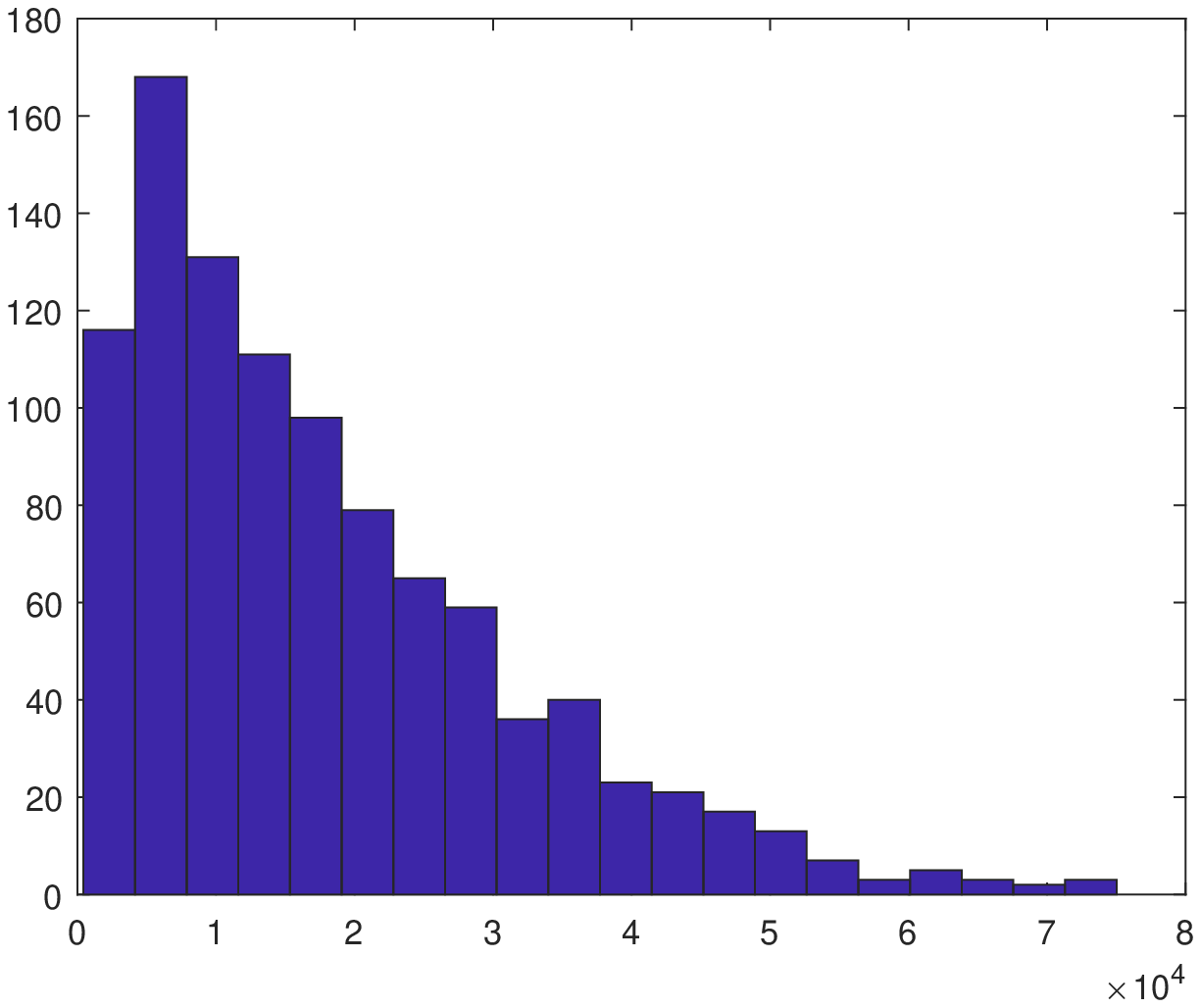}
  \includegraphics[width=6cm]{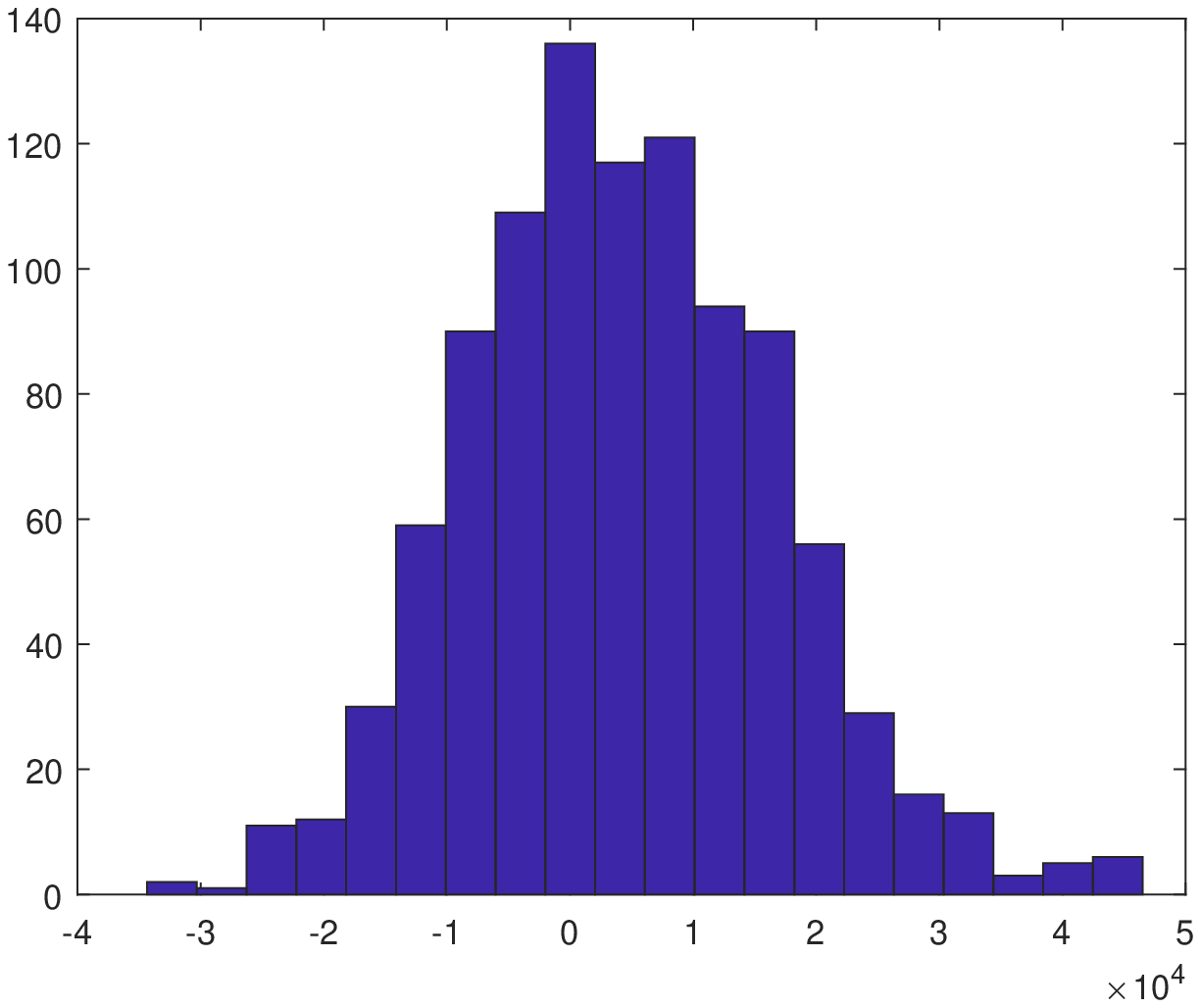}
\caption{Distributions of $U^T_{ACA}$, $U^T_{CCA}$ and their difference over the 1000 simulations done with 20-node networks.}\label{U_hist_20_node}
\end{figure}

Regarding the total resulting utility values, Fig. \ref{U_hist_20_node} depicts the distributions.

Average values of $U^T_{ACA}$ and $U^T_{CCA}$ are 13635  and 17988 in this case, while their standard deviations are 9149 and 13853 respectively.
The ACA method resulted in negative total utilities in the 5 \% of the simulated cases.
Regarding this network size, the ACA allocation method results in higher $U^T$ values in the 38.2\% of simulated cases.

Regarding the income of the auctioneer, the average total capacity payments are
23715 and 21644 in the ACA and CCA cases respectively.

The ratios of the allocated and used network capacities ($r^{ANC}$ and $r^{UNC}$) are as follows.
$$
r^{ANC}_{ACA}= 0.4918 ~~~~r^{ANC}_{CCA}=0.2753~~~~r^{UNC}_{ACA}=0.2331~~~~r^{UNC}_{CCA}=0.2753
$$

The 'unfairness' indicators ($UF$) are as
$$
UF_{ACA}=   5849.7~~~~UF_{CCA}=3947.5~~~~.
$$

The ACA method reached its final values in the 8\% of the simulated cases (in these cases no bids were submitted for the 3rd round).

\section{Discussion}
\label{sec_discussion}

One of the the most importantly required characteristics of a capacity allocation method
is the efficiency in the terms of the resulting utility of players. This corresponds to the
less formal principle of 'capacities shall be allocated to those who value them most'.
As the simulations have shown, regarding this aspect, the proposed CCA method outperforms the ACA in the majority of cases. The expected total utility of players over the analyzed high number (1000) of random scenarios was  18-41 \% higher in the case of CCA.
As the CCA method always assigns network capacities to players in a way which ensures connected paths, and is able to consider multiple alternative routes, it seems reasonable to presume that these properties of the method result in higher gains in the case of larger networks and more consumers.

If we analyze the results in the terms of allocation efficiency, we can see that the ACA method always produces a higher allocation rate ($r^{ANC}_{ACA}>r^{ANC}_{CCA}$), but the capacities allocated this way can be only partially utilized by the players (see the $r^{UNC}_{ACA}$ values and their relation to the $r^{ANC}_{ACA}$ values). In contrast, if we consider the utilized allocated capacities, the CCA method performs better in every case ($r^{UNC}_{CCA}>r^{UNC}_{ACA}$).

Regarding the income of the auctioneer, the ACA method results in 8.73 - 13.85 \% higher values. This shows that (based on the detailed simulation results) the CCA method is not the best choice if one aims to maximize the auction incomes.
Let us however emphasize that in the case of ACA, the capacity payments partially correspond to products, which are of no use in the final evaluation for the player.

Regarding the maximal difference between the maximal and the minimal utility among players, simulation results show that the CCA method results in a more fair allocation.

In addition to the evaluation of the above quantitative measures, let us point out some more differences between the two analyzed methods.

\begin{itemize}
  \item In the proposed CCA framework the players place their bid for route-quantity pairs, in contrast to the ACA framework, where bids are placed on single capacities.
  \item The ACA framework has usually multiple rounds (in the simulations, motivated by the reality of the practical applications, 3 rounds were considered), and in each round a significant number of steps are present. This means that players do have to recalculate their evaluations in each of these steps. In contrast, in the CCA framework, players evaluate their respective route-quantity pairs only once, at the beginning of the auction, which allocates capacities and determines payments in one single step.
  \item As optimal bidding is not trivial under the ACA framework (see subsection \ref{optimal_bidding_ACA}), in other words, we may say that the computational burden is put to the participants in the ACA case, where the clearing algorithm is simple, and in contrast,
      in the CCA framework bidding is simple, and the clearing algorithm is more complex.
\end{itemize}

\section{Conclusions and future work}
\label{sec_conclusions}

In this work we proposed a convex combinatorial auction method (CCA) for the allocation of capacities in capacity-constrained networks, where prior given transfer and source costs apply and the evaluation of routes is subjective by the players. We compared the proposed method with a 3-round ACA allocation method, which aims to model the current practice of capacity allocation of natural gas networks in the EU.
We performed simulations on random networks, assuming only consumers as participants of the auction, to evaluate the characteristics of these allocation methods. We found that while the ACA method in average gives higher incomes for the auctioneer, the proposed ACA method in the majority of cases results in higher and more fair resulting utility for the participants, and contributes to the more efficient utilization of network resources.

\subsection{Future wwork}

The explanation of the obtained results include some plausible, but still hypothetical assumptions (e.g. the ACA framework performs worse if the average length of the transport routes is higher), which have to be validated in further computational studies.
In addition, regarding the most important question, the possible future availability of the method, several other studies have to be performed as well.

First of all, the current modelling studies considered only consumers as participants of the auction, which is unrealistic. It is possible that under different assumptions, the results will be significantly different. If we consider a setup where consumers may bid for routes, but they are ready to take the gas at their 'doorstep'  as well (for reasonable price), the behavior of producers wishing to market their gas for maximal profit may be also included in the model. In this case, producers will also be present among participants bidding for routes/capacities and more general results will be obtainable.

Second, as we have seen, the CCA method does not always provide better utility results for players.
It is straightforward to ask how the efficiency of the CCA method depends on network topology and parameters, and more importantly, how would these methods perform on more or less detailed models of the Eurasian gas network, the structure of which is known, and its transfer and demand parameters are also known to some extent.

Although the introduction of the CCA method was the first step in the process of this analysis, several more computational studies will be required to properly characterize the practical applicability of the CCA method. In particular, the shortcomings of the European capacity allocation methods might stem from other aspects as well in addition to the intrinsic properties of the capacity allocation method used. We must note, that changing the allocation mechanism might not be the easiest and most convenient way for addressing the shortcomings.

In addition, we must note that the congestion problem seems an issue of limited significance at the European gas networks nowadays, and might be even less of an issue as gas consumption might fall due to decarbonization goals.

\section{Acknowledgments}
The author thanks Borb\'{a}la Tak\'{a}csn\'{e} T\'{o}th and P\'{e}ter Kotek for their valuable input and for the fruitful discussions. The author also thanks Tim Roughgarden who suggested to use the VCG algorithm for the problem.
This work has been supported by the Funds PD 123900, K 131 545 
of the Hungarian National Research, Development and Innovation Office. D\'{a}vid Csercsik is a grantee of the Bolyai scholarship program of the Hungarian Academy of Sciences.


\newpage
\section*{Appendix A: The 3 rounds of ACA for the proposed simple example}

\subsection*{Round 1}

As mentioned in the main text, in the initial step of the first round, all capacities of the network are subject to the auction. This means that the initial vector of capacities under auction ($CaU(1)$) may be written
as

\begin{small}
\begin{align}
\label{CaU_1_init}
CaU(1) ~=~\left[ 80   ~~ 75 ~~   70  ~~  60 ~~   60   ~~ 60 ~~   80 ~~   75   ~~ 70  ~~  60 ~~   60   ~~  60\right]
\end{align}
\end{small}

The initial price for every product is 0. In the first step, each player calculates its optimal potential flows. As at this point none of the players has any allocated capacity, straightforwardly, the $f_{aac}$ vectors for each player will be qual to 0. On the other hand, according to the concept proposed in subsection \ref{optimal_bidding_ACA}, the $f_{cau}$ vectors will be as

\begin{small}
\begin{align}
\left(
  \begin{array}{ccc}
    f_{cua}^1 & f_{cua}^2 & f_{cua}^3 \\
  \end{array}
\right)=
\left(
  \begin{array}{ccc}
    0 &          0 &          0  \\
     0 &          0 &          0  \\
     0 &          0 &          0  \\
    10 &          0 &          0  \\
     0 &          0 &          0  \\
     0 &          0 &          0  \\
    80 &         10 &          0  \\
    10 &         75 &         15  \\
     0 &          0 &         70  \\
     0 &         10 &          0  \\
     0 &          0 &          0  \\
     0 &          0 &         15  \\
  \end{array}
\right)
\end{align}
\end{small}

This will determine the bid quantities submitted ($SBQ$) in the first step

\begin{small}
\begin{align}
\label{appendix_sbq_1}
SBQ(1)=\left(
  \begin{array}{cccccccccccc}
 0 &          0 &          0 &         10 &          0 &          0 &         80 &         10 &          0 &          0 &          0 &          0  \\
    0 &          0 &          0 &          0 &          0 &          0 &         10 &         75 &          0 &         10 &          0 &          0  \\
    0 &          0 &          0 &          0 &          0 &          0 &          0 &         15 &         70 &          0 &          0 &         15  \\
  \end{array}
\right)
\end{align}
\end{small}
where the first row corresponds to the bid quantities submitted by player 1, etc.

The next step is to calculate the sum of columns of (\ref{appendix_sbq_1}), and determine the set of
products, for which there is overbidding and the set for which there is underbidding.
If we compare the bid sums
\begin{small}
$$
\sum SBQ(1) =[0~~0~~0~~10~~0~~0   ~~90~~100~~70~~10~~0~~15]
$$
\end{small}
with the capacity volumes under auction, now equal to the maximal edge capacities detailed in table
\ref{Table_edge_par_EX1}, we find that there is overbidding for line 1 and 2 in the negative direction (columns 7 and 8). For these products, the price will be increased by 1 unit in the next step, while for the others, the auction ends, and capacities are allocated to the bidders at their actual price.
Thus, the matrix of allocated capacities ($AC$) after the first auction step will be

\begin{small}
\begin{align}
\label{appendix_AC_1s1}
AC(1)=\left(
  \begin{array}{cccccccccccc}
  0 &          0 &          0 &         10 &          0 &          0 &          0 &          0 &          0 &          0 &          0 &          0  \\
    0 &          0 &          0 &          0 &          0 &          0 &          0 &          0 &          0 &         10 &          0 &          0  \\
    0 &          0 &          0 &          0 &          0 &          0 &          0 &          0 &         70 &          0 &          0 &         15  \\
  \end{array}
\right)
\end{align}
\end{small}
where, similarly to $SBQ$, row correspond to players. As players received these capacities at their initial price (0), this step implies no capacity payments for the players.

In step 2, the only remaining products are the negative directional capacities of lines 1 and 2 (80 and 75 units respectively), at the price of 1 units. If we redo the calculations for the optimal potential flows, considering the new prices, we get the same result as before -- in other words, we may conclude that the previously determined flows are still desired by the participants, also in the case of higher capacity prices of the overbidded lines. This means that they will submit the same bids for these products in this step as before.

If we continue this line of calculations, we find that that in step 4, after recalculating optimal potential flows, player 2 drops its bid corresponding to line 1 (as the potential flow on it no longer benefits him/her at the price of 3 for the line), thus the corresponding product will be allocated to player 1 entirely at the price of 3, implying 240 units of capacity payment for player 1.

Similarly, in step 6, player 1 also drops its bid corresponding to the negative direction of line 2, but considering the bids of player 2 and 3, the overbidding remains (75+15>75). This situation remains unresolved until step 9, when the recalculation of optimal potential flows results in the following decision: Player 2 reduces its bid from 75 to 40. This can be explained by the fact, that the increase of total cost (capacity + transfer + inlet payments) implies that a flow of 75 units on line 2 is no longer profitable for him/her, but a flow of 40 is still is. The reason is that the quantity of the first step of the inverse demand curve depicted in subsection \ref{consumer_demand} is equal to 40 units, and at this total price, this is the only remaining demand part for which is worth to import the gas at this price.

This way the overbidding of line 2 in the negative direction resolves, and the resulting allocated capacities of round one of the ACA may be written as

\begin{small}
\begin{align}
\label{appendix_AC_1s8}
AC(8)=\left(
  \begin{array}{cccccccccccc}
0 &          0 &          0 &         10 &          0 &          0 &         80 &          0 &          0 &          0 &          0 &          0  \\
    0 &          0 &          0 &          0 &          0 &          0 &          0 &         40 &          0 &         10 &          0 &          0  \\
    0 &          0 &          0 &          0 &          0 &          0 &          0 &         15 &         70 &          0 &          0 &         15  \\
  \end{array}
\right)~~~~.
\end{align}
\end{small}

We may also summarize the final prices of the line capacities as

\begin{small}
$$
[0 ~~  0 ~~ 0 ~~ 0 ~~ 0 ~~ 0 ~~~~ 3 ~~ 8 ~~ 0 ~~ 0 ~~ 0 ~~ 0]
$$
\end{small}

Considering the allocated amounts and the resulting prices above, the resulting capacity payments in the first round are $240,~320$ and $120$ for players 1,2 and 3 respectively.

Next we calculate the actual optimal flows regarding the above allocation.

\begin{small}
\begin{align}
\left(
  \begin{array}{ccc}
    f_{aac}^1 & f_{aac}^2 & f_{aac}^3 \\
  \end{array}
\right)=
\left(
  \begin{array}{ccc}
     0 &          0 &          0  \\
     0 &          0 &          0  \\
     0 &          0 &          0  \\
     0 &          0 &          0  \\
     0 &          0 &          0  \\
     0 &          0 &          0  \\
    80 &          0 &          0  \\
     0 &         40 &         15  \\
     0 &          0 &         70  \\
     0 &          0 &          0  \\
     0 &          0 &          0  \\
     0 &          0 &         15  \\
  \end{array}
\right)
\end{align}
\end{small}

\begin{figure}[h!]
  \centering
  \includegraphics[width=5cm]{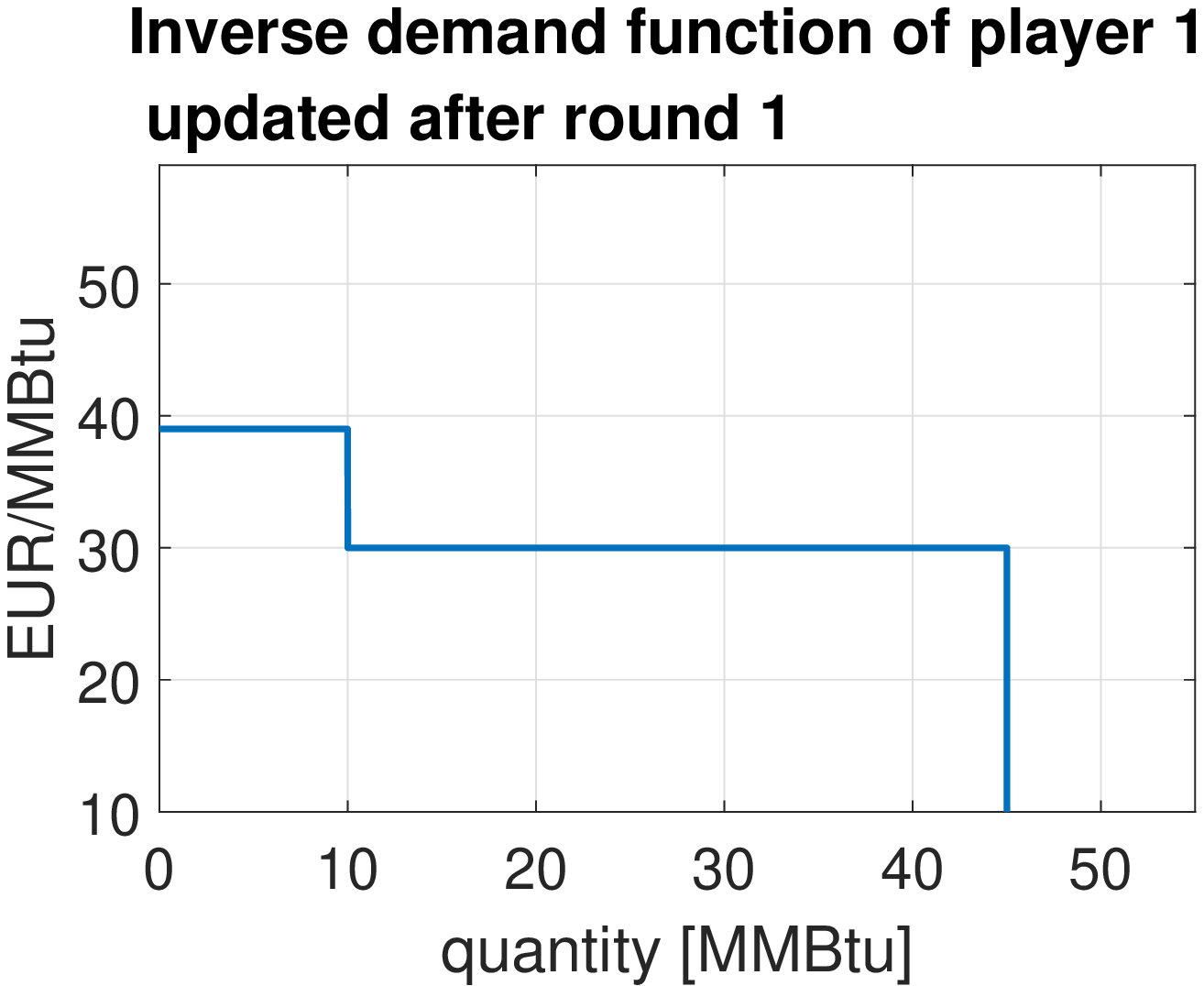}~
  \includegraphics[width=5cm]{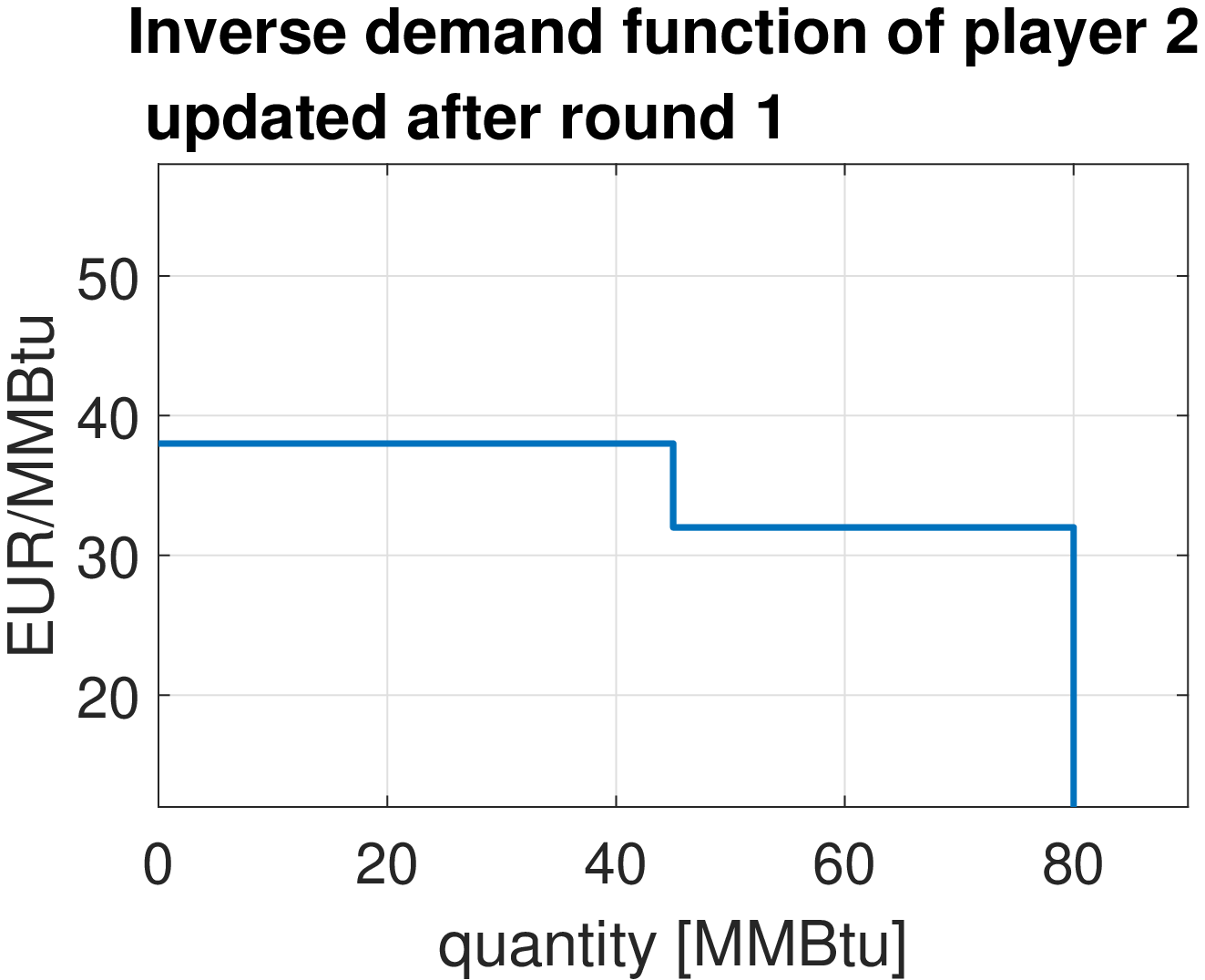}~
  \includegraphics[width=5cm]{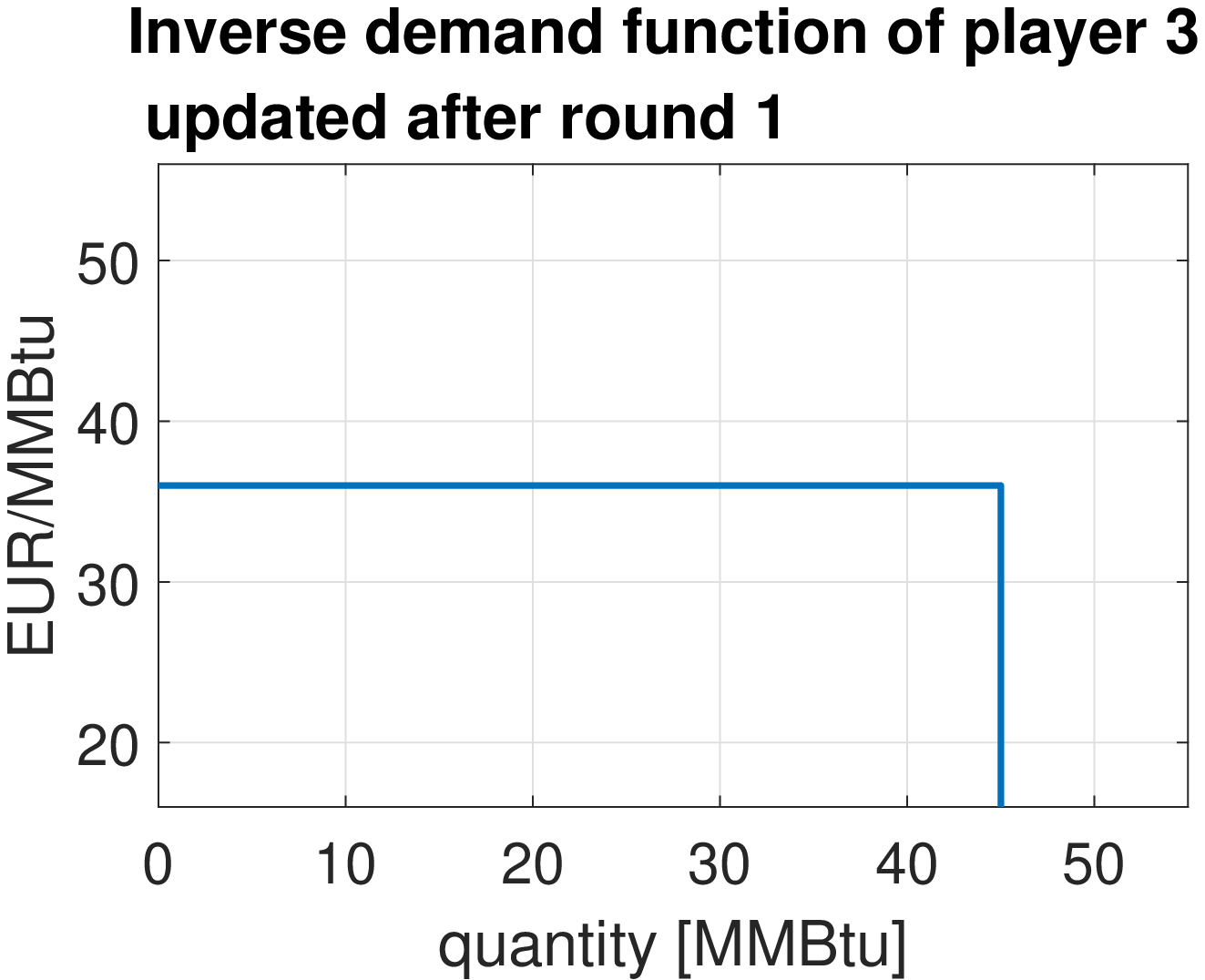}
  \caption{Updated inverse demand functions after the first round of ACA.}\label{NCA_example_DC_U1}
\end{figure}

We can see that the accessible quantities ensured after round 1 are $80,~40$ and $85$ for player 1,2 and 3 respectively (the flow of 15 units in the case of player 3 includes two edges, while the remaining flows are one-edge transfers). According to these quantities, the inverse demand functions are updated as depicted in Fig. \ref{NCA_example_DC_U1}. The updated demand curves will be used in the following ACA round.

After the first round, player 1 has 10 units of unused capacity on line 4 in the positive direction and player 2 has 10 units of unused capacity on line 4 in the negative direction.

\subsection*{Round 2}

Following the first round, the second round is initialized via the determination of the
capacities under auction. This is done by subtracting the column sums of matrix (\ref{appendix_AC_1s8}) from the vector (\ref{CaU_1_init}).

\begin{small}
\begin{align}
\label{CaU_2_init}
CaU(2)=CaU(1)-\sum AC(8) ~=~\left[ 80 ~~75 ~~70~~50~~60~~60~~~~ 0 ~~20~~0~~50~~60~~45\right]
\end{align}
\end{small}

In addition all prices are reset to 0.

Before the first step of the second round, players determine their optimal potential flows. IOn this case this results in the quantities summarized in eq. (\ref{opt_pot_flows_R2_init}).

\begin{small}
\begin{align}
\label{opt_pot_flows_R2_init}
\left(
  \begin{array}{cccccc}
    f_{aac}^1 & f_{aac}^2 & f_{aac}^3 & f_{cua}^1 & f_{cua}^2 & f_{cua}^3 \\
  \end{array}
\right)=
\left(
  \begin{array}{cccccc}
     0 &          0 &          0 &      0 &          0 &          0  \\
     0 &          0 &          0 &      0 &          0 &          0  \\
     0 &          0 &          0 &      0 &          0 &          0  \\
    10 &          0 &          0 &      0 &          0 &          0  \\
     0 &          0 &          0 &      0 &          0 &          0  \\
     0 &          0 &          0 &      0 &          0 &          0  \\
     0 &          0 &          0 &      0 &          0 &          0  \\
     0 &          0 &          0 &     10 &         20 &          0  \\
     0 &          0 &          0 &      0 &          0 &          0  \\
     0 &          0 &          0 &      0 &          0 &          0  \\
     0 &          0 &          0 &      0 &          0 &          0  \\
     0 &          0 &          0 &      0 &          0 &          0 \\
  \end{array}
\right)
\end{align}
\end{small}
It can be seen in (\ref{opt_pot_flows_R2_init}) that player 1 uses the already allocated free capacity of 10 units on edge 1 (positive direction).

Thus, the submitted bid quantities ($SBQ$) in the first step of round 2 will be

\begin{small}
\begin{align}
\label{appendix_sbq_r2s1}
SBQ(1)=\left(
  \begin{array}{cccccccccccc}
   0 &          0 &          0 &          0 &          0 &          0 &          0 &         10 &          0 &          0 &          0 &          0  \\
    0 &          0 &          0 &          0 &          0 &          0 &          0 &         20 &          0 &          0 &          0 &          0  \\
    0 &          0 &          0 &          0 &          0 &          0 &          0 &          0 &          0 &          0 &          0 &          0  \\
  \end{array}
\right)
\end{align}
\end{small}

We proceed similar to round 1. After the first step we can see that there is overbidding on product 8, namely the negative directional capacity of edge 2 (for all other products the auction ends after the first step, and no capacities are allocated). In the second step, as the price of the overbidded capacity is increased to 1, player 1 drops his/her bid, thus the capacity is allocated to player 2 at the price of 1 at the end of step 2.
The resulting allocated capacities of round two of the ACA may be written as detailed in (\ref{appendix_AC_r2s2}).

\begin{small}
\begin{align}
\label{appendix_AC_r2s2}
AC(8)=\left(
  \begin{array}{cccccccccccc}
 0 &          0 &          0 &         10 &          0 &          0 &         80 &          0 &          0 &          0 &          0 &          0  \\
    0 &          0 &          0 &          0 &          0 &          0 &          0 &         60 &          0 &         10 &          0 &          0  \\
    0 &          0 &          0 &          0 &          0 &          0 &          0 &         15 &         70 &          0 &          0 &         15  \\
  \end{array}
\right)~~~~.
\end{align}
\end{small}

Player 2 gets its capacity of 20 units at the price of 1, thus the capacity payment for him/her is 20 units in this round.

\subsection*{Round 3}
After the usual update of the $CuA$ vector and the resetting of the prices, the optimal potential flows are zero vectors for each player in this case. This means that no bids are submitted in round 3, the final results of round 2 are the final results of the ACA process in this case.

\subsection*{Evaluation of the results}

\subsubsection*{Player 1}
The total allocated capacity of Player 1 ($AC^1_T$) is

\begin{small}
\begin{align}
\label{appendix_AC_P1T}
AC_T^1=\left(
  \begin{array}{cccccccccccc}
 0 &          0 &          0 &         10 &          0 &          0 &         80 &          0 &          0 &          0 &          0 &          0  \\
  \end{array}
\right)~~~~.
\end{align}
\end{small}

As it is also naturally reflected in the optimal flow calculations, player 1 is not able to use the 10 units of capacity allocated on line 4 (+ direction), its final optimal flow $f_{ACA~F}^1$ on the available capacities is described by the vector (\ref{ACA_final_flow_P1}).

\begin{small}
\begin{align}
\label{ACA_final_flow_P1}
  \begin{array}{c}
    f_{ACA~F}^1 = [0~~0~~0~~0~~0~~0~~80~~0~~0~~0~~0~~0]^T \\
  \end{array}
\end{align}
\end{small}

The utility of consumption may be calculated from the original inverse demand function depicted in Fig. \ref{NCA_example_DC_1}: $U_C^1=50\cdot47+30\cdot39=3520$.
The optimal flow implies the transfer cost $C^1_T=720$ and the inlet cost $C^1_I=1840$. Player 1 paid a total amount of $C^1_C=250$ during the ACA process, thus its resulting utility is

\begin{equation}\label{appendix_res_utility_P1}
  U^1_{ACA}=U^1_C - C^1_T - C^1_I -C^1_C=710
\end{equation}

\subsubsection*{Player 2}
The total allocated capacity of Player 2 ($AC^2_T$) is

\begin{small}
\begin{align}
\label{appendix_AC_P2T}
AC_T^2=\left(
  \begin{array}{cccccccccccc}
    0 &          0 &          0 &          0 &          0 &          0 &          0 &         60 &          0 &         10 &          0 &          0  \\
  \end{array}
\right)~~~~.
\end{align}
\end{small}

The optimal flows of player 2 $f_{ACA~F}^2$ on the available capacities are described by the vector (\ref{ACA_final_flow_P2}).

\begin{small}
\begin{align}
\label{ACA_final_flow_P2}
  \begin{array}{c}
    f_{ACA~F}^2 = [0~~0~~0~~0~~0~~0~~0~~60~~0~~0~~0~~0]^T \\
  \end{array}
\end{align}
\end{small}

Considering that player 2 paid 330 units as capacity payment in the first round, and 40 in the second, its resulting utility may be calculated as

\begin{equation}\label{appendix_res_utility_P2}
  U^2_{ACA}=U^2_C - C^2_T - C^2_I -C^2_C= 2600   - 480 - 1380  - 370=370
\end{equation}

\subsubsection*{Player 3}
The total allocated capacity of Player 3 ($AC^3_T$) is

\begin{small}
\begin{align}
\label{appendix_AC_P3T}
AC_T^3=\left(
  \begin{array}{cccccccccccc}
    0 &          0 &          0 &          0 &          0 &          0 &          0 &         15 &         70 &          0 &          0 &         15  \\
  \end{array}
\right)~~~~.
\end{align}
\end{small}

The optimal flows of player 3 $f_{ACA~F}^3$ on the available capacities are described by the vector (\ref{ACA_final_flow_P3}).

\begin{small}
\begin{align}
\label{ACA_final_flow_P3}
  \begin{array}{c}
    f_{ACA~F}^3 = [0~~0~~0~~0~~0~~0~~0~~15~~70~~0~~0~~15]^T \\
  \end{array}
\end{align}
\end{small}

The resulting utility of player 3 may be calculated as

\begin{equation}\label{appendix_res_utility_P3}
  U^3_{ACA}=U^3_C - C^3_T - C^3_I -C^3_C= 4365    - 965  - 1955  - 205 =1240
\end{equation}

\section*{Appendix B: CCA calculation details for the proposed simple example}

\subsection*{Derivation of CCA bids}

In this subsection, we show how the parameters of the network and of the inverse demand functions clearly define a set of bids according to the format described in subsection \ref{Bid format}.

For the derivation of the plausible bid set in the case of the proposed example, let us first consider the
value of route 1 of player 1 ($R^1_1$) as an example in the case of the amount of 50 MMBtu. The source cost is 23, while the cost of edge 1 (the only edge of $R^1_1$ is 9), resulting in the total cost of 32 for route $R^1_1$, which means a cost of 1600 in the case of 50 units.

On the other hand, considering the inverse demand function of consumer 1, depicted in Fig. \ref{NCA_example_DC_1} we can see that 50 units of gas results in the consumption utility of 2350 units (=47 $\cdot$ 50). Thus we may conclude that importing 50 units of gas via $R^1_1$ produces a value of 2350-1600=750 units for player 1. As in the current study we assume truthful bidding, we assume that player 1 bid its true evaluation (750 units) for the route-quantity pair of $(R^1_1,50)$. Let us emphasize at this point that Players do not pay the nominal bidding value for their accepted bids in the CCA framework (see subsection \ref{subsection_VCG}) -- if player 1 had its bid fully accepted, and he/she would pay 750 units for the capacity rights, its resulting utility would be 0.

Let us now consider the same route in the case of 90 MMBtu (corresponding to the second step of the inverse demand function of player 1). The total cost of the transfer and the source may be derived similarly (90$\cdot$32=2880), while the resulting utility is 50 $\cdot$ 47 + 40 $\cdot$ 39= 3910 in this case. Thus importing 70 units of gas via $R^1_1$ can be evaluated to 1030 EUR by player 1.
Regarding the third step of the inverse demand function, it can be easily calculated that importing 95 units is not profitable via this route.

Based on the above calculations, let us thus assume that player 1 submits these two bids for its first route:

\begin{equation}\label{Example_1_P1_bids}
  B^1_{1,1}=(50,~750)~~~~B^1_{1,2}=(90,~1030)
\end{equation}

What happens if both bids are partially accepted? As we stated before, the outcome of the auction assigns a convex combination of the submitted bids to each player, so this is a plausible scenario.

If only the first bid is partially accepted ($0<x^1_{1,1}<1$), the case is trivial -- both the price paid for the bid and the utility implied by the transport are multiplied by $x^1_{1,1}$, which is acceptable to player 1 (the net utility will be still 0 if the nominal price is paid).

If the second bid is partially accepted ($0<x^1_{1,2}<1$), the situation may be even better.
Let us e.g. consider $x^1_{1,2}=0.6$, meaning the import of 54 units via route $R^1_{1,2}$. The nominal payment of Player 1 (not the realized payment, as emphasized before) for the route license is $0.6 \cdot 1030 = 618$ units in this case. Importing 54 units implies a cost of $54 \cdot 32=1728$, and the utility of $50 \cdot 47 +4 \cdot 39 = 2506$ which leaves player one with a surplus of net utility of 160 units (=2506-(618+1728)).

This simple calculation demonstrates that because of the non-increasing inverse demand functions, partial acceptance of bids corresponding to the second or further steps of demand curves may be only beneficial for the players.

Similar to the evaluation of $R^1_{1}$ all other routes of the players may be evaluated, and for each step of the respective demand function, bids may be derived.

We consider only those bids, for which it is true that even its full acceptance is still acceptable for the player (as we have seen this is not necessarily true for all the three steps of the inverse demand function in every case).
The numbers of bids corresponding to various players and routes are summarized in the Table \ref{table_number_of_bids}.
\begin{table}[h!]
\begin{center}
\begin{tabular}{|c|c|c|c|c|c|}
  \hline
  player & $R^i_1$ & $R^i_2$ & $R^i_3$ & $R^i_4$ & $R^i_5$ \\ \hline
  1  & 2  &   2  &   2  &   1  &   1  \\
  2  & 3  &   2  &   1  &   1  &   -  \\
  3  & 3  &   2  &   3  &   2  &   -  \\
  \hline
\end{tabular}
\caption{The number of bids corresponding to various routes of players in the proposed example.} \label{table_number_of_bids}
\end{center}
\end{table}

The quantities and prices of the submitted bids are summarized in Tables
\ref{quantities_of_bids_ex} and \ref{price_of_bids_ex} respectively.

\begin{table}[h!]
\begin{small}
\begin{center}
\begin{tabular}{|c|c|c|c|}
  \hline
  route ($j$) & k=1 & k=2 & k=3 \\ \hline
 1 &     50 &         90 &          0  \\
 2 &     50 &         90 &          0  \\
 3 &     50 &         90 &          0  \\
 4 &     50 &          0 &          0  \\
 5 &     50 &          0 &          0  \\
  \hline
\end{tabular}
~~
\begin{tabular}{|c|c|c|c|}
  \hline
  route ($j$) & k=1 & k=2 & k=3 \\ \hline
   1& 40 &         85 &        120  \\
   2& 40 &         85 &          0  \\
   3& 40 &          0 &          0  \\
   4& 40 &          0 &          0  \\
  \hline
\end{tabular}
~~
\begin{tabular}{|c|c|c|c|}
  \hline
  route ($j$) & k=1 & k=2 & k=3 \\ \hline
 1&   50 &         85 &        130  \\
 2&   50 &         85 &          0  \\
 3&   50 &         85 &        130  \\
 4&   50 &         85 &          0  \\
  \hline
\end{tabular}
\caption{The quantities of bids ($q^i_{j,k}$) corresponding to players 1, 2 and 3 in the proposed example.} \label{quantities_of_bids_ex}
\end{center}
\end{small}
\end{table}


\begin{table}[h!]
\begin{small}
\begin{center}
\begin{tabular}{|c|c|c|c|}
  \hline
  route ($j$) & k=1 & k=2 & k=3 \\ \hline
   1& 750 &       1030 &          0  \\
   2& 600 &        760 &          0  \\
   3& 425 &        445 &          0  \\
   4& 325 &          0 &          0  \\
   5& 200 &          0 &          0  \\
  \hline
\end{tabular}
\begin{tabular}{|c|c|c|c|}
  \hline
  route ($j$) & k=1 & k=2 & k=3 \\ \hline
  1&  600 &        915 &        950  \\
  2&  400 &        490 &          0  \\
  3&  140 &          0 &          0  \\
  4&  280 &          0 &          0  \\
  \hline
\end{tabular}
~~
\begin{tabular}{|c|c|c|c|}
  \hline
  route ($j$) & k=1 & k=2 & k=3 \\ \hline
1&    950 &       1475 &       1565  \\
2&    825 &       1263 &          0  \\
3&    850 &       1305 &       1305  \\
4&    675 &       1008 &          0  \\
  \hline
\end{tabular}
\caption{The prices of bids ($p^i_{j,k}$) corresponding to players 1, 2 and 3 in the proposed example.} \label{price_of_bids_ex}
\end{center}
\end{small}
\end{table}
%

\bibliographystyle{plain}
\bibliography{NCA}

\begin{thebibliography}{10}

\bibitem{bilgin2009geopolitics}
Mert Bilgin.
\newblock Geopolitics of european natural gas demand: Supplies from russia,
  caspian and the middle east.
\newblock {\em Energy Policy}, 37(11):4482--4492, 2009.

\bibitem{eu-1775-2005}
European Comission.
\newblock Council regulation ({EU}) no 1775/2005 of the european parliment and
  of the council of 28 september 2005 on conditions for access to the natural
  gas transmission networks.
\newblock {\em Official Journal of the European Union}, L289/1, 2005.

\bibitem{CSERCSIK_2019EE}
D{\'a}vid Csercsik, Franz Hubert, Bal{\'a}zs~R. Sziklai, and
  L{\'a}szl{\'o}~{\'A}. K{\'o}czy.
\newblock Modeling transfer profits as externalities in a cooperative
  game-theoretic model of natural gas networks.
\newblock {\em Energy Economics}, 80:355 -- 365, 2019.

\bibitem{ACER2016}
{David Merino}.
\newblock Implementation monitoring report on the capacity allocation
  mechanisms network code, 2016.
\newblock {Agency for the Cooperation of Energy Regulators}.

\bibitem{erdHos1960evolution}
Paul Erd{\H{o}}s and Alfr{\'e}d R{\'e}nyi.
\newblock On the evolution of random graphs.
\newblock {\em Publ. Math. Inst. Hung. Acad. Sci}, 5(1):17--60, 1960.

\bibitem{ericson2009eurasian}
Richard~E Ericson.
\newblock Eurasian natural gas pipelines: the political economy of network
  interdependence.
\newblock {\em Eurasian Geography and Economics}, 50(1):28--57, 2009.

\bibitem{eu1991directive}
EU.
\newblock {Council Directive 91/296/EEC on the transit of natural gas through
  grids}.
\newblock {\em Official Journal of the European Communities}, L147(37), 1991.

\bibitem{eu1998directive}
EU.
\newblock Directive 98/30/{EC} of the {E}uropean {P}arliament and of the
  {C}ouncil of 22 {J}une 1998 concerning common rules for the internal market
  in natural gas.
\newblock {\em Official Journal of the European Communities}, L204, 1998.

\bibitem{eu2003directive}
EU.
\newblock {Directive 2003/55/EC of the European Parliament and of the Council
  of 26 June 2003 concerning common rules for the internal market in natural
  gas and repealing Directive 98/30/EC}.
\newblock {\em Official Journal of the European Union}, L176, 2003.

\bibitem{EU2009directive_gas}
EU.
\newblock { Directive 2009/73/EC of the European Parliament and of the Council
  of 13 July 2009 concerning common rules for the internal market in natural
  gas and repealing Directive 2003/55/EC}.
\newblock {\em Official Journal of the European Union}, L211:94--136, 2009.

\bibitem{EU2009reg}
EU.
\newblock {Regulation (EC) No 715/2009 of the European Parliament and of the
  Council on conditions for access to the natural gas transmission networks and
  repealing Regulation (EC) No 1775/2005}.
\newblock {\em Official Journal of the European Union}, L211(36), 2009.

\bibitem{EU2017-CommRegulation}
EU.
\newblock Commission regulation {(EU)} 2017/459 of 16 march 2017 establishing a
  network code on capacity allocation mechanisms in gas transmission systems
  and repealing regulation ({EU}) no 984/2013.
\newblock {\em Official Journal of the European Union}, L72, 2017.

\bibitem{groves1973incentives}
Theodore Groves et~al.
\newblock Incentives in teams.
\newblock {\em Econometrica}, 41(4):617--631, 1973.

\bibitem{keller2019mergers}
Jann~T Keller, Gerard~H Kuper, and Machiel Mulder.
\newblock Mergers of germany's natural gas market areas: Is transmission
  capacity booked efficiently?
\newblock {\em Utilities Policy}, 56:104--119, 2019.

\bibitem{madani2017revisiting}
Mehdi Madani.
\newblock Revisiting european day-ahead electricity market auctions: {MIP}
  models and algorithms.
\newblock {\em PhD Thesis, Universite Catholique de Louvain}, 2017.

\bibitem{scherr1975pricing}
Bruce~A Scherr and Emerson~M Babb.
\newblock Pricing public goods: An experiment with two proposed pricing
  systems.
\newblock {\em Public Choice}, pages 35--48, 1975.

\bibitem{REKK2017}
Borb\'{a}la Tak\'{a}csn\'{e}~T\'{o}th, P\'{e}ter Kotek, and P\'{e}ter
  Kaderj\'{a}k.
\newblock {Evaluating the 6 March PRISMA capacity auctions}.
\newblock
  \url{https://rekk.hu/article/145/evaluating-the-6-march-prisma-capacity-auctions},
  2017.
\newblock {Regional Centre for Energy Policy Research}.

\bibitem{vickrey1961counterspeculation}
William Vickrey.
\newblock Counterspeculation, auctions, and competitive sealed tenders.
\newblock {\em The Journal of finance}, 16(1):8--37, 1961.

\end{thebibliography}

\end{document}